\documentclass[preprint,showpacs,preprintnumbers,amsmath,amssymb]{revtex4}

\usepackage{graphicx}
\usepackage{epsfig}
\usepackage{amsmath}
\usepackage{graphics}

\begin{document}

\title{Transmission of Excitations in a Spin-1 \\ Bose--Einstein Condensate through a Barrier}

\author{Shohei Watabe$^{1}$} 
\altaffiliation{Present Address: 
Department of Physics, Keio University, 3-14-1 Hiyoshi, Kohoku-ku, Yokohama 223-8522, Japan}
\altaffiliation{CREST(JST), 4-1-8 Honcho, Kawaguchi, Saitama 332-0012, Japan}
\author{Yusuke Kato$^{2}$}
\affiliation{
$^{1}$
Department of Physics, The University of Tokyo, Tokyo 113-0033, Japan}
\affiliation{
$^{2}$
Department of Basic Science, The University of Tokyo, Tokyo 153-8902, Japan}

\begin{abstract} 
We investigate tunneling of excitations across a potential barrier separating two spin-1 Bose--Einstein condensates. 
Using mean-field theory at absolute zero temperature, 
we determine the transmission coefficients of excitations 
in the saturated magnetization state and unsaturated magnetization states. 
All excitations, except the quadrupolar spin mode in the saturated magnetization state, 
show the anomalous tunneling phenomenon characterized as perfect tunneling 
in the low-momentum limit through a potential barrier. 
The quadrupolar spin mode in the saturated magnetization state, whose spectrum is massive, shows total reflection. 
We discuss properties common between excitations showing the anomalous tunneling phenomenon. 
Excitations showing perfect tunneling have a gapless spectrum in the absence of the magnetic field, 
and their wave functions in the low-energy limit are the same as the condensate wave function. 
\end{abstract}

\maketitle

\section{Introduction}\label{ICT}

Almost ten years ago, an interesting property of an interacting Bose--Einstein condensed system was predicted: 
a potential barrier is fully penetrable for an excitation of a Bose--Einstein condensate (BEC) in the low-energy limit~\cite{Kovrizhin2001}. 
Since a potential barrier generally inhibits transmission of a particle and 
leads to total reflection in the low-momentum limit, 
this perfect transmission in the BEC, termed anomalous tunneling~\cite{Kagan2003}, 
has attracted much attention and has been studied extensively 
and intensively~\cite{Kagan2003,Danshita2006,FujitaMThesis,FujitaUnpublished,Kato2007,Ohashi2008,Watabe2008,Watabe2009RefleRefra,Tsuchiya2008,Takahashi2009,WatabeKato2009,Takahashi2010,Tsuchiya2009,WatabeKatoLett}. 

The following crucial facts for the anomalous tunneling of the excitation in a BEC (i.e., the Bogoliubov excitation) have been discussed: 
the fact that the wave function of the Bogoliubov excitation has the same form as the condensate wave function in the low-momentum limit~\cite{Kato2007} and the fact that the wave function of its excitation with small but finite momentum behaves as the condensate one with a small supercurrent~\cite{Ohashi2008}, 
and so on. 

However, these facts were found through studies of a specific system (i.e., a scalar BEC) 
so that discussions and understandings of anomalous tunneling are restrictive.
If we can find another system showing anomalous tunneling, 
it will lead to a deeper understanding of anomalous tunneling. 
An efficient approach in this direction is to examine tunneling phenomena in 
a BEC of particles with spin-1 degrees of freedom (the so called spin-1 spinor BEC). 

The spin-1 spinor BEC of dilute atomic gases was realized in 1998~\cite{Stamper-Kurn1998,Stenger1998}. 
In the same year, Ohmi and Machida~\cite{Ohmi1998} and also Ho~\cite{Ho1998} determined ground states and excitations in the spin-1 spinor BEC 
through mean-field theory. 
In this BEC, there are two phases for ground states, a ferromagnetic phase and a polar phase, each of which has three-types of excitations. 
For instance, the ferromagnetic phase has the Bogoliubov mode whose energy $E$ is linear in the momentum $|{\bf p}|$, 
the transverse spin mode with $E \propto {\bf p}^{2}$, and the quadrupolar spin wave whose energy is massive.

The following questions will be answered through studies of the spin-1 spinor BEC:
Is the linear dispersion essential for anomalous tunneling of excitations? 
(Does an excitation with $E \propto {\bf p}^{2}$ tunnel through the potential barrier without reflection in the limit $|{\bf p}| \rightarrow 0$?) 
Do excitations that exhibit anomalous tunneling always have gapless spectra? 
If we can deduce the essence of anomalous tunneling from the answers to these questions, 
then it will useful for predicting tunneling properties in other systems. 

\begin{figure}[tbp]
\begin{center}
\includegraphics[width=7cm]{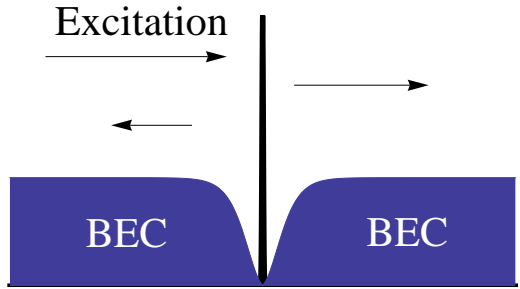}
\end{center}
\caption{(Color online) 
A schematic picture of the tunneling problem of excitations in a Bose-Einstein condensate against a potential barrier. 
}
\label{fig1}
\end{figure}

In this paper, we thus investigate tunneling problems of excitations in the spin-1 spinor BEC. 
An example of this problem is schematically shown in Fig.~\ref{fig1}. 
An excitation runs toward a potential barrier from one side 
and is transmitted through and reflected against the barrier. 
Excitations are scattered off not only the potential barrier but the condensate wave function deformed by the potential barrier. 
The density of the BEC is depleted only around the potential barrier because of the nonlinear effect 
due to the interparticle interaction. 
We study both the ferromagnetic and the polar phases, but we focus rather on the saturated magnetization state $|{\bf m}|=1$, 
and an unsaturated magnetization state $0\leq |{\bf m}|<1$, 
which are, respectively, associated with the ferromagnetic and the polar phases in the absence of the magnetic field. 
Here, $|{\bf m}|$ is the normalized magnetization, where the magnetization is fully polarized when $|{\bf m}|=1$. 
As well as transmission coefficients, 
we examine wave functions of excitations in the low-energy limit 
to study a characteristic specific to excitations showing anomalous tunneling. 

Table~\ref{tableI} summarizes our main results in this paper. 
We find that all excitations but the quadrupolar spin mode show anomalous tunneling 
in the low-momentum limit. 
We also find that these excitations have the following two features: 
(i) the energy gap is zero in the absence of external fields 
and (ii) the wave functions of excitations in the low-momentum limit have the same forms as condensate wave functions. 

\begin{table}[tbp]
\begin{center}
\caption{
Transmission coefficients $T$ in the low-momentum limit $|{\bf p}| \rightarrow 0$. }
{Saturated magnetization state ($|{\bf m}|=1$)}    
\\        
\begin{tabular}{ccccccc}
\hline
\hline
\\[-10pt]
\shortstack{Bogoliubov excitation } && $\lim\limits_{|{\bf p}|\rightarrow 0}T = 1$ \\
\shortstack{transverse spin excitation } && $\lim\limits_{|{\bf p}|\rightarrow 0}T = 1$ \\
\shortstack{quadrupolar spin excitation } && $\lim\limits_{|{\bf p}| \rightarrow 0}T = 0$ \\
\hline
\hline
\end{tabular}
\\[15pt]
Unsaturated magnetization state ($0\leq |{\bf m}|<1$)
\\
\begin{tabular}{ccccccc}
\hline
\hline
\\[-10pt]
 \shortstack{Bogoliubov and longitudinal \\ spin excitation (Out-of-Phase) } &&&  \shortstack{ $\lim\limits_{|{\bf p}|\rightarrow 0}T  = 1 $ \\[-3pt]} \\[5pt] 
 
 \shortstack{Bogoliubov and longitudinal \\ spin excitation (In-Phase) } &&& \shortstack{ $\lim\limits_{|{\bf p}|\rightarrow 0}T  = 1 $ \\[-3pt]} \\[5pt] 
 
 \shortstack{transverse spin excitation} &&& $\lim\limits_{|{\bf p}|\rightarrow 0}T  = 1$ \\ 
\hline
\hline
\end{tabular}
\end{center}
\label{tableI}
\end{table} 

Using a $\delta$-function potential barrier, 
we explicitly extract differences between the transverse spin mode and the quadrupolar spin mode in the saturated magnetization state, 
both of which have spectra with a term proportional to ${\bf p}^{2}$. 
It is worthwhile to compare these two modes through an exactly solvable model, 
since it gives a clue toward understanding how perfect tunneling or perfect reflection is induced. 
A scalar BEC is presumably the simplest system in the sense that it has no internal degrees of freedom. 
However we note that the transverse spin wave of the spin-1 spinor BEC 
is the simplest mode showing anomalous tunneling, rather than the Bogoliubov mode in the scalar BEC; the transverse spin wave is described by a single-component Schr\"odinger-type equation while the Bogoliubov mode obeys a two-component Schr\"odinger-type equation. 

Through our studies, the features of anomalous tunneling of the Bogoliubov mode in a scalar BEC will be clear. 
Our results will also lead to further understandings of spin excitations in spinor BECs. 
Moreover, the tunneling problem of excitations is an important issue in ferromagnets~\cite{Demokritov2004,Hansen2007}. 
Beyond the field of cold atoms, 
the present study leads to a further understanding of the spin-wave tunneling in ferromagnets.

This paper is organized as follows. 
We review the mean-field theory of the spin-1 BEC in Sec.~\ref{SecII}, where the Gross-Pitaevskii-type equation and the phase diagram of the ground state are derived. 
The Wronskian and the energy flux are also discussed to determine the transmission coefficients. 
Section~\ref{SecIII} presents the tunneling properties of excitations in the saturated magnetization state. 
Section~\ref{SecIV} presents the tunneling properties of excitations in an unsaturated magnetization state. 
In Sec.~\ref{SecV}, we investigate the tunneling problems of excitations across the $\delta$-function potential barrier 
in the saturated magnetization state. 
Using an exact solution of the wave function, we derive the transmission coefficient of the transverse spin mode. 
We also analytically discuss perfect reflection of the quadrupolar spin mode using the $\delta$-function potential barrier. 
In Sec.~\ref{SecVI}, we examine the dependence of the tunneling properties on the spin-dependent interaction strength and the linear Zeeman effect. 
Appendix~\ref{AppendixA} serves as a summary of the eigenmodes 
and Appendix~\ref{AppendixC} presents the tunneling properties of the unsaturated magnetization state in the integrable case.

\section{Mean-field theory of spin-1 BEC}\label{SecII}

\subsection{Gross-Pitaevskii-type equation and ground state}

We summarize the mean-field theory of a weakly interacting spin-1 Bose system developed by Ohmi and Machida~\cite{Ohmi1998} and Ho~\cite{Ho1998} in order to fix notations.
We start with a Hamiltonian of a spin-1 Bose system described by 
\begin{align}
\hat{H} = &
\int d{\bf  r}\sum\limits_{i}
\hat{\Psi}_{i}^{\dag} ({\bf r}) 
\left [ 
-\frac{\hbar^{2}}{2m} \nabla^{2}
+ V_{\rm ext} ({\bf r}) 
\right ]
\hat{\Psi}_{i}({\bf r})
\nonumber 
\\
&
+\frac{c_{0}}{2}
\int d{\bf r} 
\sum\limits_{i, j }
\hat{\Psi}_{i}^{\dag} ({\bf r}) 
\hat{\Psi}_{j}^{\dag} ({\bf r})
\hat{\Psi}_{j} ({\bf r}) 
\hat{\Psi}_{i} ({\bf r})
\nonumber
\\ 
&
+\frac{c_{1}}{2}
\int d{\bf r} 
\sum\limits_{k }
\sum\limits_{i, j, i', j'  }
\hat{\Psi}_{i}^{\dag} ({\bf r}) 
\hat{\Psi}_{i'}^{\dag} ({\bf r})
S_{ij}^{k} S_{i' j'}^{k}
\hat{\Psi}_{j'} ({\bf r})
\hat{\Psi}_{j} ({\bf r}) 
\nonumber
\\ 
&
-g\mu_{\rm B}B 
\int d{\bf r} \sum\limits_{ij}
\hat{\Psi}_{i}^{\dag} ({\bf r})
S_{ij}^{z}
\hat{\Psi}_{j} ({\bf r}), 
\label{TBSpB1}
\end{align} 
where the sums with respect to $i$, $j$, $i'$ and $j'$ 
are taken over the magnetic sublevels $\pm 1$ and $0$. 
$\hat{\Psi}_{i}({\bf r})$ is the Bose field operator satisfying $[\hat{\Psi}_{i}({\bf r}), \hat{\Psi}_{j}^{\dag}({\bf r}')] = \delta_{ij}\delta({\bf r}-{\bf r}')$, 
$[\hat{\Psi}_{i}({\bf r}), \hat{\Psi}_{j}({\bf r}')] = 0$, 
and $[\hat{\Psi}_{i}^{\dag}({\bf r}), \hat{\Psi}_{j}^{\dag}({\bf r}')] = 0$. 
The sum with respect to $k$ is taken over the Cartesian coordinates $x$, $y$, and $z$; 
$m$ is the mass, $g$ is the Land\'e's $g$ factor, $\mu_{\rm B}$ is the Bohr magneton, 
and $B (\geq 0)$ is the strength of a magnetic field (with the $z$ axis being taken as the direction of the magnetic field). 
The interaction strengths are given by $c_{0} = 4\pi\hbar^{2}(2a_{2} + a_{0})/(3m)$, and 
$c_{1}= 4\pi\hbar^{2}(a_{2} - a_{0})/(3m)$, 
where $a_{S_{\rm tot}}$ is the $s$-wave scattering length for a collision process with total spin $S_{\rm tot}$. 
The spin matrices of the spin-1 system are given by 
\begin{align}
S^{x} = 
\frac{1}{\sqrt{2}}
\begin{pmatrix}
0&1&0 \\
1&0&1 \\
0&1&0
\end{pmatrix}, 
S^{y} = 
\frac{i}{\sqrt{2}}
\begin{pmatrix}
0&-1&0 \\
1&0&-1 \\
0&1&0
\end{pmatrix}, 
S^{z} = 
\begin{pmatrix}
1&0&0 \\
0&0&0 \\
0&0&-1
\end{pmatrix}.  
\label{TBSpB2}
\end{align}

The field operator obeys the equation of motion $i\hbar \partial \hat{\Psi}_{j}/\partial t = [\hat{\Psi}_{j}, \hat{H}]$. 
Within the mean-field theory $\langle \hat{\Psi}_{j} \rangle = \Phi_{j}e^{-i\mu_{j} t/\hbar}$, 
the Gross--Pitaevskii(GP)-type equation of the spin-1 BEC 
 for stationary states is given by 
\begin{eqnarray}
\left ( \hat{\mathcal H}  \mp g \mu_{\rm B} B \right  ) \Phi_{\pm 1} 
+ c_{1}( F_{\mp} \Phi_{0}/\sqrt{2} \pm F_{z}\Phi_{\pm1}) = 0, 
\label{TBSpB5}
\\
\hat{\mathcal H}  \Phi_{0} 
+ c_{1}(F_{+} \Phi_{+1} + F_{-}\Phi_{-1})/\sqrt{2} = 0, 
\label{TBSpB6}
\end{eqnarray}
where $\hat{\mathcal H}  = -\hbar^{2}\nabla^{2}/(2m) + V_{\rm ext} -\mu + c_{0} n$ 
and $F_{+} \equiv F_{-}^{*} \equiv \sqrt{2} {\rm Re}[(\Phi_{+1}^{*}\Phi_{0} + \Phi_{0}^{*}\Phi_{-1})]$. 
We define also $n \equiv |\Phi_{+1}|^{2} + |\Phi_{0}|^{2} + |\Phi_{-1}|^{2}$ and $F_{z} \equiv |\Phi_{+1}|^{2} - |\Phi_{-1}|^{2}$. 
We use $\mu_{j} = \mu$ for $j = \pm1$ and $0$, satisfying the condition $2\mu_{0} = \mu_{+1} + \mu_{-1}$~\cite{Nistazakis2007}.

In the following, we consider the phase diagram of the ground state in spatially uniform systems in the presence of magnetic fields. It is easy to see that a fully polarized state
\begin{align}
( \Phi_{+1}, \Phi_{0}, \Phi_{-1})^{\rm T}
= 
(\sqrt{n},  0, 0)^{\rm T} 
\label{TBSpB17}
\end{align}
is a solution to (\ref{TBSpB5}) and (\ref{TBSpB6}). 
The total energy per unit volume 
$
{\mathcal E} = - \mu n + c_{0}n^{2}/2 + {\mathcal E}', 
$
with
\begin{align}
{\mathcal E}' \equiv 
c_{1}
\left [
\Phi_{+1}\Phi_{-1}\Phi_{0}^{* 2} + \Phi_{+1}^{*}\Phi_{-1}^{*}\Phi_{0}^{2}
+ 
|\Phi_{0}|^{2} \left ( |\Phi_{+1}|^{2} + |\Phi_{-1}|^{2}  \right )
\right ] 
\nonumber 
\\
+ \frac{c_{1}}{2}
(|\Phi_{+1}|^{2} - |\Phi_{-1}|^{2} )^{2}
-g\mu_{\rm B}B 
(|\Phi_{+1}|^{2}-|\Phi_{-1}|^{2}), 
\label{TBSpB12}
\end{align} 
for (\ref{TBSpB17}) is given by ${\mathcal E}' = c_{1}n^{2}/2 - g\mu_{\rm B} B$. 
A state may be characterized by the magnetization ${\bf M} = (M^x,M^y,M^z)$ with $M^k= \mu_{\rm B}\sum_{j=\pm1, 0}\Phi^*_i S_{ij}^k \Phi_j$ for $k=x,y,z$. 
When we denote by ${\bf m}$ the normalized magnetization [i.e., ${\bf m} \equiv {\bf M}/(\mu_{\rm B} n)$], 
$|{\bf m}|=1$ follows for (\ref{TBSpB17}). 
We name the state (\ref{TBSpB17}) with $|{\bf m}|=1$ the saturated magnetization state. 
Another solution to (\ref{TBSpB5}) and (\ref{TBSpB6}) is given by 
\begin{align}
( \Phi_{+1}, \Phi_{0}, \Phi_{-1})^{\rm T} 
= 
(0, \sqrt{n} , 0)^{\rm T}, 
\label{TBSpB14}
\end{align} 
which has ${\mathcal E}' = 0$ and ${\bf m}=0$. 
When $(0 \leq) g\mu_{\rm B}B < c_{1}n$, 
the state
\begin{align}
( \Phi_{+1}, \Phi_{0}, \Phi_{-1})^{\rm T}
= 
\sqrt{n/2}
(\pm\sqrt{1+A}, 0, \sqrt{1-A})^{\rm T}, 
\label{TBSpB16}
\end{align} 
with $A=g\mu_{\rm B}B/c_{1}n$ is the third solution to (\ref{TBSpB5}) and (\ref{TBSpB6}). This solution has  ${\mathcal E}' = - (g\mu_{\rm B} B)^{2}/(2c_{1}n)$ and $|{\bf m}| = A$. We name a state with $|{\bf m}|\in [0,1)$ an unsaturated magnetization state. 
Among (\ref{TBSpB17}), (\ref{TBSpB14}) and (\ref{TBSpB16}), the solution with the lowest ${\cal E}'$ yields the ground state.  
The ground state is given by the saturated magnetization state (\ref{TBSpB17})
when $c_1 n <g\mu_{\rm B}B$, while 
the state (\ref{TBSpB16}) is realized as the ground state when $(0 \leq) g\mu_{\rm B}B < c_{1}n$.

When $B = 0$ and $c_{1} < 0$, 
the saturated magnetization state having $|{\bf m}| = 1$ 
is called the ferromagnetic phase~\cite{Ho1998}. 
When $B=0$ and $c_1>0$, however, the unsaturated magnetization state having $|{\bf m}| = 0$ 
is called the polar phase~\cite{Ho1998}. 
Note that the degeneracy of the ground state relating to the gauge symmetry and the spin rotational symmetry exists when $B = 0$. 
A configuration of an order parameter 
(the condensate wave function)
is transformed into another through the gauge transformation $e^{i\theta_{0}}$ 
and the spin rotations ${\mathcal U}(\alpha, \beta, \gamma) = e^{-i S^{z}\alpha}e^{-i S^{y}\beta}e^{-i S^{z}\gamma}$. 
In the ferromagnetic phase, 
the order parameter is given by~\cite{Ho1998} 
\begin{align}
& ( \Phi_{+1}, \Phi_{0}, \Phi_{-1})^{\rm T}
\nonumber 
\\
= & 
e^{i\theta_{0}}{\mathcal U}(\alpha, \beta, \gamma)
(\sqrt{n}, 0, 0) ^{\rm T}
\nonumber 
\\ 
= & 
\sqrt{n}e^{i(\theta_{0}-\gamma)}
\left ( e^{-i\alpha} \cos^{2}\frac{\beta}{2}, 
\sqrt{2}\cos\frac{\beta}{2} \sin\frac{\beta}{2}, 
e^{i\alpha} \sin^{2}\frac{\beta}{2} \right)^{\rm T}. 
\end{align} 
In the polar phase, 
the order parameter is given by~\cite{Ho1998} 
\begin{align}
& ( \Phi_{+1}, \Phi_{0}, \Phi_{-1})^{\rm T}
\nonumber 
\\
= & 
e^{i\theta_{0}}{\mathcal U}(\alpha, \beta, \gamma)
( 0, \sqrt{n}, 0)^{\rm T}
\nonumber 
\\
= &
\sqrt{n}e^{i\theta_{0}}
\left ( 
-\frac{1}{\sqrt{2}} e^{-i\alpha} \sin\beta, 
\cos\beta, 
\frac{1}{\sqrt{2}} e^{i\alpha}\sin\beta
\right )^{\rm T}.   
\end{align} 
When one takes the $z$ axis as the spin-quantization axis~\cite{Ho1998}, 
a configuration of the order parameter in the polar phase is reduced to $(\Phi_{+1}, \Phi_{0}, \Phi_{-1})^{\rm T} = (0,\sqrt{n},0)^{\rm T}$. 
The spin quantization axis for $(\Phi_{+1}, \Phi_{0}, \Phi_{-1})^{\rm T} = (-\sqrt{n/2},0,\sqrt{n/2})^{\rm T}$ 
is the $x$ axis, and for $(\Phi_{+1}, \Phi_{0}, \Phi_{-1})^{\rm T} = (\sqrt{n/2},0,\sqrt{n/2})^{\rm T}$, 
it is the $y$ axis. 
As a result, these configurations are equivalent, and the state (\ref{TBSpB14}) is equivalent to the state (\ref{TBSpB16}) when $B=0$.

\subsection{Equation of excitations and transmission coefficient}\label{ChapSpin1}

In order to study small oscillations of the system around equilibrium, 
we consider a small fluctuation $\tilde{\phi}_{i}({\bf r}, t)$ deviated from the condensate wave function $\Phi_{i}({\bf r})$, 
where $\tilde {\phi}_{i}({\bf r}, t)$ is regarded as the classical field ($c$-number). 
The equation of excitations is given by 
\begin{align}
i \hbar \frac{\partial \tilde{\phi}_{\pm 1}}{\partial t} 
= &  
\left [ \hat{\mathcal H} + R_{\pm 1,\pm 1}'^{(+)} + c_{1} ( \pm  F_{z} + |\Phi_{0}|^{2})  \mp g \mu_{\rm B} B 
 \right ]   \tilde{\phi}_{\pm 1} 
\nonumber 
\\
&
+ 
R_{\pm1,\pm1}^{(+)} \tilde{\phi}_{\pm 1}^{*} 
+ 
P_{\pm 1}\tilde{\phi}_{0} 
+ 
R_{0,\pm1}^{(+)}  \tilde{\phi}_{0}^{*} 
\nonumber 
\\
&
+ R_{\mp 1,\pm 1}'^{(-)} \tilde{\phi}_{\mp 1} 
+ 
( R_{+1,-1}^{(-)}  + c_{1} \Phi_{0}^{2} ) \tilde{\phi}_{\mp 1}^{*}, 
\label{TBSpB7}
\\
i \hbar \frac{\partial \tilde{\phi}_{0}}{\partial t} 
= & 
\left ( \hat{\mathcal H} 
+ c_{1} n  + R_{0,0}'^{(-)}
\right ) \tilde{\phi}_{0} 
+ ( c_{0} \Phi_{0}^{2} + 2c_{1} \Phi_{+1} \Phi_{-1} ) \tilde{\phi}_{0}^{*} 
\nonumber 
\\
&
+ \sum\limits_{m=\pm 1} 
(P_{m}^{*} \tilde{\phi}_{m} + R_{0,m}^{(+)} \tilde{\phi}_{m}^{*} ), 
\label{TBSpB8}
\end{align}
where $P_{\pm 1} \equiv  ( c_{0} + c_{1} ) \Phi_{0}^{*}\Phi_{\pm 1} + 2 c_{1}  \Phi_{\mp 1}^{*}\Phi_{0}$, 
$R_{i,j}^{(\pm)} \equiv (c_{0}\pm c_{1})\Phi_{i} \Phi_{j}$, and $R_{i,j}'^{(\pm)} \equiv (c_{0}\pm c_{1})\Phi_{i}^{*} \Phi_{j}$.

A spatially constant quantity is important to determine the transmission and reflection coefficients of excitations. 
We derive this quantity from Eqs. (\ref{TBSpB7}) and (\ref{TBSpB8}). 
By assuming the Bogoliubov-type wave function $\tilde{\phi}_{i}({\bf r},t)= \sum\limits_{l}[u_{i}^{l}({\bf r}) e^{-iE_{l} t/\hbar} - (v_{i}^{l}({\bf r}))^{*} e^{iE_{l} t/\hbar}]$, 
with  a label $l$ of an eigenmode, 
we can obtain the Bogoliubov-type equation for the spin-1 BEC from Eqs. (\ref{TBSpB7}) and (\ref{TBSpB8}). 
With the use of this equation, we have the following relation: 
\begin{align}
&
(E^{l'} - E^{l})\sum\limits_{i = 0, \pm 1} 
[   u_{i}^{l'} ({\bf r}) ( u_{i}^{l} ({\bf r}) )^{*}  -   v_{i}^{l'} ({\bf r}) ( v_{i}^{l}({\bf r}) )^{*}    ]
\nonumber
\\ 
= & 
-\frac{\hbar^{2}}{2m} \nabla \cdot
{\bf W}_{l,l'} ({\bf r}), 
\end{align}
where the Wronskian ${\bf W}_{l,l'} ({\bf r})$ is given by 
\begin{align}
{\bf W}_{l,l'} ({\bf r})  \equiv & 
\sum\limits_{i = 0, \pm1}
\left [ 
(u_{i}^{l'}({\bf r}))^{*}  \nabla u_{i}^{l}({\bf r})  - u_{i}^{l} ({\bf r})  \nabla (u_{i}^{l'}  ({\bf r}))^{*}
\right . 
\nonumber
\\
& + 
\left .
(v_{i}^{l'}({\bf r}))^{*}   \nabla v_{i}^{l}({\bf r})   - v_{i}^{l}  ({\bf r}) \nabla (v_{i}^{l'}  ({\bf r}))^{*}
\right ] .
\end{align} 
Here $l$ and $l'$ denote labels of eigenmodes. 
When eigenmodes $l$ and $l'$ have the same energy $E \equiv E^{l}= E^{l'}$, 
one finds ${\bf W}_{l,l'} ({\bf r})= {\rm const.}$ 
Hereafter, we discuss the Wronskian ${\bf W}_{l,l} $ (i.e., $l' = l$) to determine the transmission and reflection coefficients. 

This Wronskian of the Bogoliubov equation is related to the time-averaged energy flux~\cite{Kagan2003,Danshita2006}. 
An equation of the 
total energy ${\cal E}$ per a unit volume
 is given by 
\begin{align} 
\frac{\partial {\cal E}}{\partial t} 
= 
- \nabla 
\cdot 
 {\bf Q}, 
\end{align}
where the energy flux ${\bf Q}$ is defined as 
\begin{align} 
{\bf Q} 
\equiv 
- 
\frac{\hbar^{2}}{2m} 
\sum\limits_{i = 0, \pm1} 
{\rm Re} 
\left [ 
\frac{\partial  {\Psi}_{i}^{*}} {\partial t} 
\nabla {\Psi_{i}}
\right ]. 
\end{align} 
Let us introduce a time-averaged quantity $\langle X \rangle$ 
given by 
\begin{align}
\langle X  \rangle = \frac{1}{\tau} \int_{0}^{\tau} X dt,  
\end{align} 
where $\tau$ is the time given by $\tau = 2\pi \hbar n /E$ ($n$ is integer) and 
$E$ is the excitation energy. 

We consider the time averaged energy flux of the eigenmode $l$, 
assuming that the small fluctuation deviated from the order parameter $\Phi_{j}({\bf r})$ is given by the Bogoliubov form, i.e., 
\begin{align} 
\Psi_{j} = 
\Phi_{j} + 
u_{j}^{l} e^{-iE_{l} t/\hbar} - (v_{j}^{l})^{*} e^{iE_{l} t/\hbar}.  
\end{align} 
As a result, the time averaged energy flux $\langle {\bf Q}_{l}\rangle$ of the eigenmode $l$ is given by
\begin{align} 
\langle {\bf Q}_{l}\rangle
= & - 
\frac{i \hbar E_l}{4m} 
{\bf W}_{l,l}. 
\end{align} 
Since ${\bf W}_{l,l}$ is spatially constant, 
the time-averaged energy flux of an eigenmode $l$ is also independent of the position. 

We give the transmission and reflection coefficients using this spatially constant quantity. 
In this paper, we study one-dimensional tunneling problems, assuming the normal incidence of excitations against a potential wall.   
We name the incident side of the tunneling problem the left side and label it as L. 
We label the right side as R. 
The spin-1 spinor BEC has some types of excitation because of the internal degrees of freedom. 
In particular, as mentioned below (and also in Appendix~\ref{AppendixA}), 
three modes in the uniform system are well known~\cite{Ohmi1998,Ho1998}, 
which may characterize the incident, reflected, and transmitted waves. 
Let us suppose that the incident mode is one of these three excitations in the spin-1 BEC, 
which we name $\sigma = {\rm I}$. We name the other two modes $\sigma = {\rm II}$ and ${\rm III}$~\cite{NoteMode}. 
A wave function far from the potential barrier on the left side 
is described by a sum of the incident and reflected waves: 
\begin{align}
& 
( 
u_{+1,{\rm L}}^{l} ( {\bf r} ), 
v_{+1,{\rm L}}^{l} ( {\bf r} ), 
u_{0,{\rm L}}^{l} ( {\bf r} ), 
v_{0,{\rm L}}^{l} ( {\bf r} ), 
u_{-1,{\rm L}}^{l} ( {\bf r} ), 
v_{-1,{\rm L}}^{l} ( {\bf r} )
)^{\rm T}
\nonumber 
\\
\rightarrow & 
(e^{i k_{\rm I} x} + r_{\rm I}e^{-i k_{\rm I} x} )
\times 
( 
\tilde{u}_{+1,{\rm L}}^{\rm I} , 
\tilde{v}_{+1,{\rm L}}^{\rm I} ,
\tilde{u}_{0,{\rm L}}^{\rm I} ,
\tilde{v}_{0,{\rm L}}^{\rm I} ,
\tilde{u}_{-1,{\rm L}}^{\rm I} ,
\tilde{v}_{-1,{\rm L}}^{\rm I} 
)^{\rm T}
\nonumber
\\
& 
+ 
\sum\limits_{\sigma = {\rm II}, {\rm III}}
r_{\sigma}e^{-i k_{\sigma} x } 
\times 
( 
\tilde{u}_{+1,{\rm L}}^{\sigma} , 
\tilde{v}_{+1,{\rm L}}^{\sigma} ,
\tilde{u}_{0,{\rm L}}^{\sigma} ,
\tilde{v}_{0,{\rm L}}^{\sigma} ,
\tilde{u}_{-1,{\rm L}}^{\sigma} ,
\tilde{v}_{-1,{\rm L}}^{\sigma} 
)^{\rm T}
.  
\end{align}
The wave function far from the potential barrier on the right side is described by the transmitted wave: 
\begin{align}
& 
( 
u_{+1,{\rm R}}^{l} ( {\bf r} ), 
v_{+1,{\rm R}}^{l} ( {\bf r} ), 
u_{0,{\rm R}}^{l} ( {\bf r} ), 
v_{0,{\rm R}}^{l} ( {\bf r} ), 
u_{-1,{\rm R}}^{l} ( {\bf r} ), 
v_{-1,{\rm R}}^{l} ( {\bf r} )
) ^{\rm T}
\nonumber
\\
\rightarrow & 
\sum\limits_{\sigma = {\rm I}, {\rm II}, {\rm III}}
t_{\sigma} e^{i k_{\sigma} x} 
\times 
( 
\tilde{u}_{+1,{\rm R}}^{\sigma} , 
\tilde{v}_{+1,{\rm R}}^{\sigma} ,
\tilde{u}_{0,{\rm R}}^{\sigma} ,
\tilde{v}_{0,{\rm R}}^{\sigma} ,
\tilde{u}_{-1,{\rm R}}^{\sigma} ,
\tilde{v}_{-1,{\rm R}}^{\sigma} 
)^{\rm T}. 
\end{align}
Here, $r_{\sigma}$ and $t_{\sigma}$, respectively, are the amplitude reflection and transmission coefficients of a mode $\sigma$. 
The time-averaged energy flux of a mode $\sigma$ in each asymptotic regime is given by 
\begin{align} 
\langle Q_{{\rm I}, \rm L}\rangle
=& 
+ \frac{\hbar E}{2m} 
k_{{\rm I},{\rm L}} (1-|r_{{\rm I}}|^{2})
\sum\limits_{i = 0, \pm 1} (|u_{i, {\rm L}}^{{\rm I}}|^{2} +|v_{i, {\rm L}}^{{\rm I}}|^{2} ), 
\\
\langle Q_{\sigma, \rm L}\rangle
=& 
- \frac{\hbar E}{2m} 
k_{\sigma,{\rm L}} |r_{\sigma}|^{2} 
\sum\limits_{i = 0, \pm 1} (|u_{i, {\rm L}}^{\sigma}|^{2} +|v_{i, {\rm L}}^{\sigma}|^{2} ), \quad ({\rm for} \, \sigma = {\rm II \, and \, III}),  
\\
\langle Q_{\sigma, \rm R}\rangle
=& 
+ \frac{\hbar E}{2m} 
k_{\sigma,{\rm R}} |t_{\sigma}|^{2}
\sum\limits_{i = 0, \pm 1} (|u_{i, {\rm R}}^{\sigma}|^{2} +|v_{i, {\rm R}}^{\sigma}|^{2} ), \quad ({\rm for} \, \sigma = {\rm I, II \, and \, III}). 
\end{align} 

The transmission and reflection coefficients $T_{\rm \sigma}$ and $R_{\rm \sigma}$ of a mode $\sigma$ are, respectively, defined by 
the ratios of transmitted energy flux to incident energy flux, 
and also the ratios of reflected energy flux to incident energy flux. 
Using this definition, we obtain $T_{\rm \sigma}$ and $R_{\rm \sigma}$ as 
\begin{align}
T_{\sigma} = &
\frac{
k_{\sigma,{\rm R}} 
\sum\limits_{i = 0, \pm1} (|u_{i,{\rm R}}^{{\sigma}}|^{2} +|v_{i,{\rm R}}^{{\sigma}}|^{2} )
}{
k_{{\rm I},{\rm L}}
\sum\limits_{i = 0, \pm1} (|u_{i,{\rm L}}^{{\rm I}}|^{2} +|v_{i,{\rm L}}^{{\rm I}}|^{2} ) 
} 
|t_{\sigma}|^{2}, 
\\
R_{\rm \sigma}  = & 
\frac{
k_{\sigma,{\rm L}} 
\sum\limits_{i = 0, \pm1} (|u_{i,{\rm L}}^{{\sigma}}|^{2} +|v_{i,{\rm L}}^{{\sigma}}|^{2} )
}{
k_{{\rm I},{\rm L}}
\sum\limits_{i = 0, \pm1} (|u_{i,{\rm L}}^{{\rm I}}|^{2} +|v_{i,{\rm L}}^{{\rm I}}|^{2} ) 
} 
|r_{{\sigma}}|^{2}. 
\end{align}
According to $\langle  Q_{l, \rm L}\rangle = \langle Q_{l, \rm R}\rangle$, 
one has $\sum_{\sigma} ( T_{\sigma} + R_{\sigma} ) = 1$.

These are a general framework for the tunneling problem of excitations in the spin-1 spinor BEC. 
In the next section, 
we examine the tunneling probability of excitations in the spin-1 BEC focusing on the saturated magnetization state. 
After that, we investigate the unsaturated magnetization state.

\section{saturated magnetization state}\label{SecIII}

\begin{figure}[tbp]
\begin{center}
\includegraphics[width=6cm]{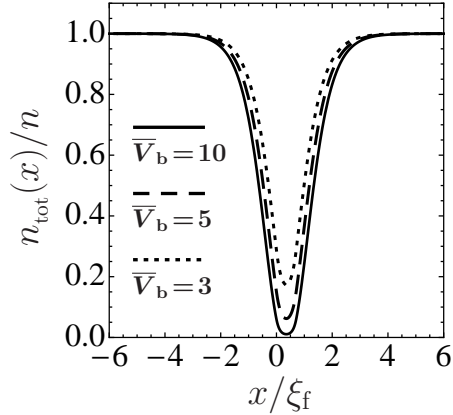}
\end{center}
\caption{
Total density profiles of the saturated magnetization state in the presence of the rectangular potential barrier (\ref{PotNumSPIN}), 
with the various potential heights 
$\overline{V}_{\rm b} \equiv V_{\rm b}/(\xi_{\rm f}c_{+}n)$ and with $V_{\rm L} = V_{\rm R} = 0$. 
}
\label{fig2}
\end{figure}

The saturated magnetization state is realized when $c_{1}n < g\mu_{\rm B} B$. 
Its condensate wave function of the uniform system is given by 
$
(
\Phi_{+1} , 
\Phi_{0} , 
\Phi_{-1}
)^{\rm T} 
= 
(
\sqrt{n} , 
0 , 
0 
)^{\rm T}$. 
Substituting the above wave function into the GP-type equation, 
one obtains the chemical potential $\mu = c_{+}n - g\mu_{\rm B}B$, where $c_{+} \equiv c_{0} + c_{1}$.  
It is expected that wave functions $\Phi_{0}$ and $\Phi_{-1}$ do not appear near a potential barrier, 
since a modulation of condensate wave functions costs kinetic energy; 
thus, we assume the following condensate wave function near the potential wall: 
\begin{align}
(
\Phi_{+1} , 
\Phi_{0} , 
\Phi_{-1}
)^{\rm T}
= 
(
\sqrt{n} \phi (x) ,
0 ,
0
)^{\rm T}. 
\label{Spin1SecFerro-CW}
\end{align} 
We have checked the validity of this assumption through the numerical calculations of the GP-type equations (\ref{TBSpB5}) and (\ref{TBSpB6}). 
Substituting (\ref{Spin1SecFerro-CW}) into the GP-type equation, 
we have 
\begin{align}
0 = 
\left (
-\frac{\hbar^{2}}{2m} \frac{d^{2}}{dx^{2}}
+ V_{\rm ext} -
c_{+} n 
\right ) \phi 
+ 
c_{+} n \phi^{3}. 
\label{satuGPP}
\end{align} 
We notice that this equation is the same as the GP equation of the scalar BEC.
Figure~\ref{fig2} shows the density profiles for the rectangular potential barrier given by 
\begin{align}
V(x) = 
\left \{ 
\begin{array}{ll}
V_{\rm L} &\qquad {\rm for} \quad x \leq 0, 
\\
V_{\rm b} &\qquad {\rm for} \quad 0 < x \leq \xi_{\rm f}/\sqrt{2}, 
\\
V_{\rm R} &\qquad {\rm for} \quad \xi_{\rm f}/\sqrt{2} < x. 
\end{array}
\right .
\label{PotNumSPIN}
\end{align} 
The healing length of the ferromagnetic state is given by $\xi_{\rm f} \equiv \hbar/\sqrt{m c_{+} n}$. 
The magnetization is proportional to this profile. 
As in the schematic of Fig.~\ref{fig1}, 
the condensate density is depleted around the potential barrier. 

Using this wave function for the condensate, 
we obtain the following equations of fluctuations: 
\begin{align}
i\hbar \frac{\partial \tilde{\phi}_{+1}}{\partial t} 
= & 
\left [
-\frac{\hbar^{2}}{2m} \frac{d^{2}}{dx^{2}}
+ V_{\rm ext} +
c_{+} n (\phi^{2}-1)
\right ] \tilde{\phi}_{+1} 
+ 
c_{+} n\phi^{2}
(\tilde{\phi}_{+1}+\tilde{\phi}_{+1}^{*}),  
\\
i\hbar \frac{\partial \tilde{\phi}_{0}}{\partial t} 
= & 
\left [
-\frac{\hbar^{2}}{2m} \frac{d^{2}}{dx^{2}}
+ V_{\rm ext} + 
c_{+} n (\phi^{2}-1)
+ 
g\mu_{\rm B}B 
\right ] \tilde{\phi}_{0}, 
\\ 
i\hbar \frac{\partial \tilde{\phi}_{-1}}{\partial t} 
= & 
\left (
-\frac{\hbar^{2}}{2m} \frac{d^{2}}{dx^{2}}
+ V_{\rm ext} 
-
c_{+} n 
+ 
c_{-} n
\phi^{2}
+ 2 g\mu_{\rm B}B 
\right ) \tilde{\phi}_{-1}, 
\end{align}
where $c_{-} \equiv c_{0} - c_{1}$. 

We use the Bogoliubov transformation 
$\tilde {\phi}_{+1} = u_{+1} \exp{(-iE t/\hbar)} - v_{+1}^{*} \exp{(iE t/\hbar)}$ 
and introduce two functions, $S_{+1} \equiv u_{+1} + v_{+1}$ and $G_{+1} \equiv u_{+1} - v_{+1}$,  for the magnetic sublevel $+1$. 
To obtain the transmission coefficient for the spin component $+1$, 
we solve the following equation: 
\begin{align}
E G_{+1} 
&= 
\left [
-\frac{\hbar^{2}}{2m} \frac{d^{2}}{dx^{2}}
+ V_{\rm ext} (x)+ 
c_{+} n (\phi^{2}-1)
\right ] S_{+1}, 
\label{FerroBogoEq1}
\\
E S_{+1} 
&= 
\left [
-\frac{\hbar^{2}}{2m} \frac{d^{2}}{dx^{2}}
+ V_{\rm ext} (x)+ 
c_{+} n (3\phi^{2}-1)
\right ] G_{+1}, 
\label{FerroBogoEq2}
\end{align}
imposing the boundary condition 
\begin{align}
\begin{pmatrix}
S_{+1} \\
G_{+1}
\end{pmatrix} 
= & 
\left [ \exp{(i k x)} + r  \exp{(- i k x)} \right ]
\begin{pmatrix}
\alpha_{+1} \\
\beta_{+1}
\end{pmatrix} 
+ 
a \exp{(\kappa x)} 
\begin{pmatrix}
\beta_{+1} \\
- \alpha_{+1}
\end{pmatrix} & 
{\rm for} \,
x \rightarrow - \infty, 
\\
\begin{pmatrix}
S_{+1} \\
G_{+1}
\end{pmatrix} 
= & 
t \exp{(i k x)} 
\begin{pmatrix}
\alpha_{+1} \\
\beta_{+1}
\end{pmatrix}
+ 
b \exp{(- \kappa x)} 
\begin{pmatrix}
\beta_{+1} \\
- \alpha_{+1}
\end{pmatrix} &
{\rm for} \, x \rightarrow + \infty, 
\end{align}
where $k$ and $\kappa$ are, respectively, given by 
$\hbar k \equiv \sqrt{2m [ \sqrt{ (c_{+} n)^{2} + E^{2}} - c_{+}n ] }$ 
and 
$\hbar \kappa \equiv \sqrt{2m [ \sqrt{(c_{+}n)^{2} + E^{2}} + c_{+} n ] }$. 
The excitation of the $+1$ component becomes the ordinary Bogoliubov excitation 
with $E = \sqrt{\varepsilon ( \varepsilon + 2 c_{+} n ) }$, 
where $\varepsilon \equiv \hbar^{2}k^{2}/(2m)$, 
as noted in Appendix~\ref{AppendixA}. $\alpha_{+1}$ and $\beta_{+1}$ are also given in Appendix~\ref{AppendixA}. 

For excitations $\tilde{\phi}_{0}$ and $\tilde{\phi}_{-1}$, 
we solve the equations 
\begin{align}
E\tilde{\phi}_{0}
& = 
\left [
-\frac{\hbar^{2}}{2m} \frac{d^{2}}{dx^{2}}
+ V_{\rm ext} + 
c_{+} n (\phi^{2}-1)
+ 
 g\mu_{\rm B}B 
\right ] \tilde{\phi}_{0}, 
\label{Spin1ExcitationFerroEq0}
\\
E\tilde{\phi}_{-1}
& = 
\left (
-\frac{\hbar^{2}}{2m} \frac{d^{2}}{dx^{2}}
+ V_{\rm ext} -
c_{+} n
+ 
c_{-} n
\phi^{2}
+ 2 g\mu_{\rm B}B 
\right ) \tilde{\phi}_{-1}, 
\label{Spin1ExcitationFerroEqMinus}
\end{align}
imposing the boundary conditions 
\begin{align} 
\tilde{\phi}_j \rightarrow & \exp{(i k x)} + r \exp{(- i k x)}   & {\rm for} \quad x \rightarrow - \infty, 
\\
\tilde{\phi}_j \rightarrow & t \exp{( i k x)} &{\rm for} \quad x \rightarrow + \infty, 
\end{align}
with $j=0$ and $-1$. Here the momentum $\hbar k$ for the spin component $0$ 
is given by $\hbar k = \sqrt{ 2m  (E - g\mu_{\rm B}B)}$ and for the spin component $-1$, 
by $\hbar k = \sqrt{ {2m} \left [ E + 2 (c_{1}n-g\mu_{\rm B}B) \right ]}$. 
Excitations of the magnetic sublevels $0$ and $-1$ are, respectively, associated with the transverse spin mode 
and the quadrupolar spin mode with $E = \varepsilon + g\mu_{\rm B}B$ and $E = \varepsilon - 2 c_{1} n + 2g\mu_{\rm B}B$, as noted in Appendix~\ref{AppendixA}. 
Only the quadrupolar spin excitation has the non-zero inherent energy gap $\Delta_{B = 0} = 2|c_{1}|n$ for $k\rightarrow 0$ when $B=0$. 
We use the term ``the inherent energy gap'' as the energy gap in the case when $B=0$. 
The other excitations in the saturated magnetization state have no inherent energy gap (i.e., $\Delta_{B= 0} = 0$).

\begin{figure}[tbp]
\begin{center}
\includegraphics[width=6cm]{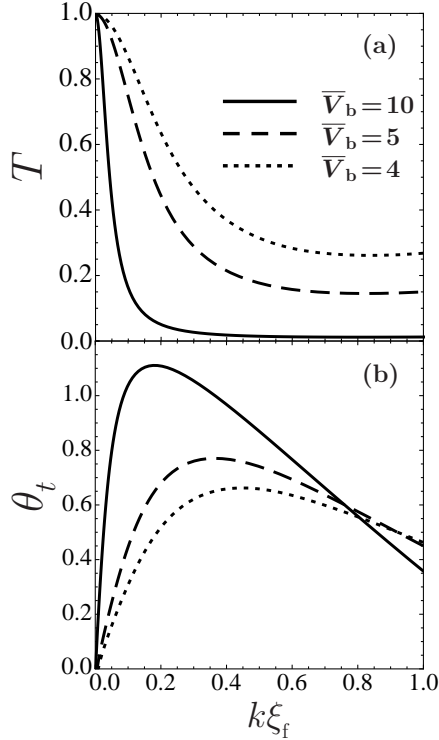}
\end{center}
\caption{
(a) Transmission coefficients $T$ and (b) corresponding phase shifts $\theta_{t}$ (defined by the argument of the amplitude transmission coefficient $t$) 
of the Bogoliubov mode with the spin component $+1$ in the saturated magnetization state 
through the rectangular barrier (\ref{PotNumSPIN}). 
We used $V_{\rm L} = V_{\rm R} = 0$, and define $\overline{V}_{\rm b} \equiv V_{\rm b}/(c_{+}n)$. 
We use parameters $a_{0} : a_{2} = 110: 107$ following Ref.~\cite{Ho1998}. 
}
\label{fig3}
\end{figure} 

\begin{figure}[tbp]
\begin{center}
\includegraphics[width=8.5cm]{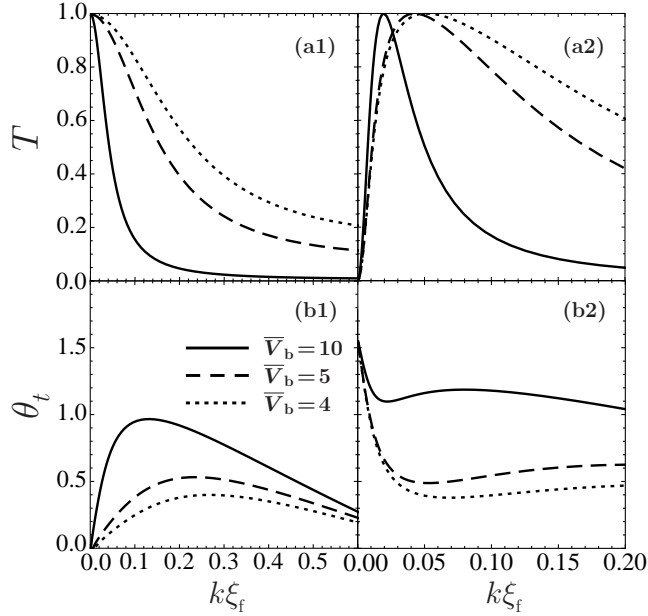}
\end{center}
\caption{ 
(a1) Transmission coefficients $T$ and (b1) corresponding phase shifts $\theta_{t}$
of the transverse spin wave mode in the saturated magnetization state. 
(a2) Transmission coefficients $T$ and (b2) corresponding phase shifts $\theta_{t}$
of the quadrupolar spin wave mode in the saturated magnetization state. 
The barrier (\ref{PotNumSPIN}) and the parameters $a_{0}$ and $a_{2}$ are the same as those in Fig.~\ref{fig3}. 
}
\label{fig4}
\end{figure}

We show the momentum dependence of the transmission coefficients and phase shifts for all three excitations 
in Figs.~\ref{fig3} and \ref{fig4}. 
We find that 
the Bogoliubov mode and the transverse spin wave show 
the anomalous tunneling phenomenon (i.e., perfect tunneling in the long-wavelength limit), although these two excitations have different spectra.  
Phase shifts, defined as the argument of the amplitude transmission coefficient $t$, of both excitations reach zero 
in the low-momentum limit. 
Unlike these two excitations, the quadrupolar spin mode does not show the anomalous tunneling phenomenon. 
Its phase shift does not reach zero in the low-momentum limit; it reaches $\pi/2$ in this case.

\begin{figure}[tbp]
\begin{center}
\includegraphics[width=8.5cm]{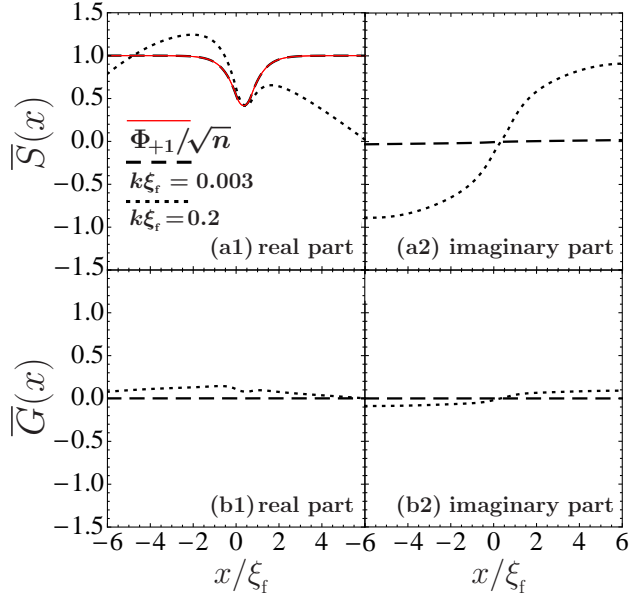}
\end{center}
\caption{(Color online) 
Wave functions of the spin component $+1$ in the saturated magnetization state. 
$\overline{S}_{}(x) \equiv {S}_{+1}(x)\sqrt{E/ ( 2 c_{+} n)}$ and $\overline{G}_{}(x) \equiv {G}_{+1}(x)\sqrt{E/( 2 c_{+} n)}$. 
We use the barrier (\ref{PotNumSPIN}), where $V_{\rm L} = V_{\rm R} = 0$ and $V_{\rm b} = 3 c_{+}n$. 
The parameters $a_{0}$ and $a_{2}$ are also the same as in Fig.~\ref{fig3}. 
}
\label{fig5}
\end{figure}

Interesting characteristics can be seen in the wave functions of the excitations. 
Figure~\ref{fig5} plots wave functions of the Bogoliubov mode in the presence of the rectangular potential barrier~(\ref{PotNumSPIN}). 
One finds that the wave function $\sqrt{E/(2c_{+}n)}{S}_{+1}(x)$ in the long wavelength regime 
coincides with the condensate wave function (i.e., $S_{+1} \propto \Phi_{+1}$). 
The existence of this type of solution in the limit $E\rightarrow 0$ can be easily confirmed by comparing the GP equation  (\ref{satuGPP}) with 
(\ref{FerroBogoEq1}). 
From Fig.~\ref{fig5}, however, one confirms that ${G}_{+1}(x)$ is absent in the long wavelength regime (i.e., $\lim_{k\rightarrow 0}G_{+1} = 0$). 
As a result, one finds that $(u_{+1},v_{+1}) \propto (\Phi_{+1},\Phi_{+1})$ follows. 
Through studies of anomalous tunneling in the scalar BEC, 
it was pointed out that total transmission in the long wavelength limit is deeply related 
to a correspondence between a wave function of the Bogoliubov excitation in the low energy limit and that of the condensate~\cite{Kato2007}. 
This characteristic for the Bogoliubov mode in the scalar BEC is 
the same as that for the Bogoliubov mode in the spin-1 BEC.

\begin{figure}[tbp]
\begin{center}
\includegraphics[width=8.5cm]{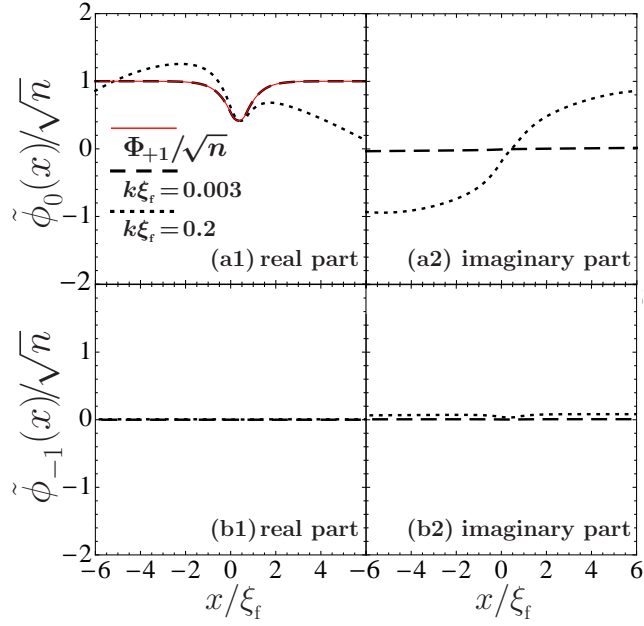}
\end{center}
\caption{(Color online) 
Spatial profiles of the wave function. 
(a1) and (a2) 
plot
$\tilde{\phi}_{0}$; (b1) and (b2) 
plot 
$\tilde{\phi}_{-1}$. 
The potential barrier and parameters are the same as those in Fig.~\ref{fig5}. 
}
\label{fig6}
\end{figure}

Let us focus on the other two modes in the saturated magnetization state: 
the transverse spin-wave mode associated with $\tilde{\phi}_{0}$, 
and the quadrupolar spin mode associated with $\tilde{\phi}_{-1}$. 
Figure~\ref{fig6} plots the wave functions of these modes. 
From panels (a1) and (a2), 
we see that 
the wave function $\tilde{\phi}_{0}$ 
 in the low-energy limit 
has the same form as the condensate wave function $\Phi_{+1}$. 
Comparing the GP equation (\ref{satuGPP}) with (\ref{Spin1ExcitationFerroEq0}) for the limit $k\rightarrow 0$, 
we can 
confirm the existence of a solution being proportional to the condensate wave function $\Phi_{+1}$ (i.e., $\tilde{\phi}_{0} \propto \Phi_{+1}$). 
As for the spin component $-1$, 
there are no solutions proportional to $\Phi_{+1}$ for $k\rightarrow 0$. 
From panels (b1) and (b2), we find that the amplitude of $\tilde{\phi}_{-1}$ goes to zero as the wavelength becomes longer; this leads to the total reflection (i.e., $t = 0$ and $r=-1$) in the limit $k\rightarrow 0$. 
This is a characteristic feature of the quadrupolar spin mode, which was noted in our earlier work~\cite{WatabeKato2009}. 
We will revisit the quadrupolar excitation in Sec.~\ref{SecV}, 
and show that the potential barrier makes the amplitude of its wave function absent in the long wavelength limit.

\section{Unsaturated magnetization state}\label{SecIV} 
\begin{figure}[tbp]
\begin{center}
\includegraphics[width=6cm]{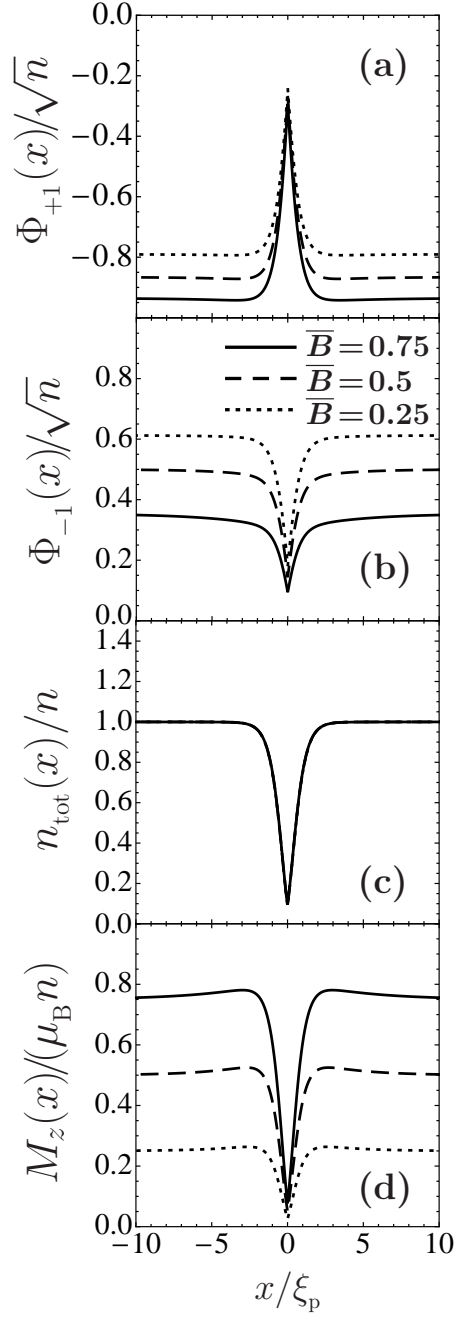}
\end{center}
\caption{(Color online) Condensate wave function (a) $\Phi_{+1}$, (b) $\Phi_{-1}$, 
(c) the local density and (d) the magnetization for various magnetic fields 
in the presence of the $\delta$-function potential barrier $V_{\rm ext}(x) = V_{\rm b}\delta(x)$ 
for $V_{\rm b}/\xi_{\rm p} = 3 c_{0}n$, 
where $\xi_{\rm p} \equiv \hbar / \sqrt{mc_{0} n}$. 
$\overline{B} \equiv g \mu_{\rm B}B/(c_{1}n)$. 
We use parameters $a_{0} : a_{2} = 46 : 52$ following Ref.~\cite{Ho1998}. 
} 
\label{fig7}
\end{figure}

We examine the transmission coefficients of excitations in the unsaturated magnetization state, 
which is realized when $(0 \leq) g\mu_{\rm B} B < c_{1}n$. 
The wave function of its ground state in the uniform system is given by Eq. (\ref{TBSpB16}). 
Substituting this configuration into the GP equation, 
the chemical potential is given by $\mu = c_{0} n$.  
$\Phi_{0}$ is expected to be absent everywhere, for the same reason mentioned in the previous section, 
and hence we assume the following configuration: 
\begin{align}
(\Phi_{+1}, \Phi_{0}, \Phi_{-1} )^{\rm T} 
= 
(\Phi_{+1}, 0, \Phi_{-1})^{\rm T} . 
\label{Eq43/2010/11/5}
\end{align} 
We have checked its validity through the numerical calculations of the GP-type equation (\ref{TBSpB5}). 
Substituting the above configuration into the GP-type equation, 
we obtain the following equation of $\Phi_{\pm 1}$: 
\begin{align}
0 & = 
\left (
-\frac{\hbar^{2}}{2m} \frac{d^{2}}{dx^{2}}
+ V_{\rm ext} -c_{0} n_{0 } \mp g\mu_{\rm B}B 
\right ) \Phi_{\pm 1}
+ 
c_{+} \Phi_{\pm 1}^{3} 
+ 
c_{-} \Phi_{\mp 1}^{2} \Phi_{\pm 1}. 
\label{GPPolarFM+1}
\end{align} 
In Fig.~\ref{fig7}, 
we show the condensate wave functions, number densities of particles, and the magnetization for various magnetic fields 
in the presence of the potential barrier. 
It is clearly seen that the density profile does not depend on the uniform magnetic field, 
while the profile of the magnetization is easily changed by the uniform magnetic field.

On the basis of the configuration (\ref{Eq43/2010/11/5}), 
we obtain equations of fluctuations, which are given by 
\begin{align}
i\hbar \frac{\partial \tilde{\phi}_{\pm 1}}{\partial t} 
& = 
\left (
-\frac{\hbar^{2}}{2m} \frac{d^{2}}{dx^{2}}
+ V_{\rm ext} 
- c_{0} n 
\mp  g\mu_{\rm B}B 
+ 
c_{+} \Phi_{\pm 1}^{2} 
+ 
c_{-} \Phi_{\mp 1}^{2} 
\right ) 
\tilde{\phi}_{\pm 1}
\nonumber
\\
& 
+ 
c_{+} \Phi_{\pm 1}^{2} 
(\tilde{\phi}_{\pm 1}  + \tilde{\phi}_{\pm 1}^{*} )
+ 
c_{-} \Phi_{+1}\Phi_{-1} 
(\tilde{\phi}_{\mp 1}  + \tilde{\phi}_{\mp 1}^{*} ), 
\label{BogoliubovPolarFMpm1}
\\
i\hbar \frac{\partial \tilde{\phi}_{0}}{\partial t} 
& = 
\left [
-\frac{\hbar^{2}}{2m} \frac{d^{2}}{dx^{2}}
+ V_{\rm ext} 
- c_{0} n 
+ 
c_{+} (\Phi_{+1}^{2} + \Phi_{-1}^{2})  
\right ] \tilde{\phi}_{0} 
+ 2c_{1} \Phi_{+1}\Phi_{-1} \tilde{\phi}_{0}^{*}. 
\end{align} 
When we introduce the Bogoliubov-type transformation $\tilde{\phi}_{j} = u_{j} e^{-iEt/\hbar} - v_{j}^{*} e^{+iEt/\hbar}$ 
and define $S_{j} \equiv u_{j} + v_{j}$ and $G_{j} \equiv u_{j} - v_{j}$, 
we obtain the following equations: 
\begin{align}
E G_{+1} 
& = 
\hat{\mathcal H}_{+}
S_{+1}, 
\label{EqSGPM48}
\\
E S_{+1} 
&= 
\left (
\hat{\mathcal H}_{+}
+ 2 c_{+} \Phi_{+1}^{2}
\right ) 
G_{+1} 
+ 2 c_{-} \Phi_{+1}\Phi_{-1} G_{-1}, 
\label{EqSGPM49}
\\
E G_{-1} 
&= 
\hat{\mathcal H}_{-}
S_{-1}, 
\label{EqSGPM50}
\\
E S_{-1} 
&= 
\left (
\hat{\mathcal H}_{-}
+ 2 c_{+} \Phi_{-1}^{2}
\right ) 
G_{-1} 
+ 2 c_{-} \Phi_{+1}\Phi_{-1} G_{+1}, 
\label{unsatuBogoPM}
\end{align}
where 
$\hat{\mathcal H}_{\pm} \equiv 
-{\displaystyle \frac{\hbar^{2}}{2m} \frac{d^{2}}{dx^{2}} } 
+ V_{\rm ext} 
-c_{0}n \mp  g\mu_{\rm B}B 
+ c_{+} \Phi_{\pm 1}^{2} + c_{-} \Phi_{\mp 1}^{2} 
$. 
In order to determine the transmission coefficients, 
we impose the following boundary condition: 
\begin{align}
\begin{pmatrix}
S_{+1} \\
G_{+1}  \\
S_{-1}  \\ 
G_{-1} 
\end{pmatrix} 
= & 
e^{ik_{(\pm)}x}
\begin{pmatrix}
\alpha_{(\pm)} \\
\beta_{(\pm)} \\
\bar{\alpha}_{(\pm)} \\
\bar{\beta}_{(\pm)}
\end{pmatrix} 
+ 
\sum\limits_{j = \pm}
r_{(j)} 
e^{-ik_{(j)}x}
\begin{pmatrix}
\alpha_{(j)} \\
\beta_{(j)} \\
\bar{\alpha}_{(j)} \\
\bar{\beta}_{(j)}
\end{pmatrix}  
+ 
\sum\limits_{j = \pm} 
a_{(j)} 
e^{\kappa_{(j)}x}
\begin{pmatrix}
-\beta_{(j)} \\
\alpha_{(j)} \\
-\bar{\beta}_{(j)} \\ 
\bar{\alpha}_{(j)}
\end{pmatrix} 
\qquad & 
{\rm for} \, x \rightarrow - \infty, 
\\
\begin{pmatrix}
S_{+1} \\
G_{+1}  \\
S_{-1}  \\ 
G_{-1} 
\end{pmatrix} 
= & 
\sum\limits_{j = \pm}
t_{(j)} 
e^{ik_{(j)}x}
\begin{pmatrix}
\alpha_{(j)} \\
\beta_{(j)} \\
\bar{\alpha}_{(j)} \\
\bar{\beta}_{(j)}
\end{pmatrix}  
+ 
\sum\limits_{j = \pm} 
b_{(j)} 
e^{-\kappa_{(j)}x}
\begin{pmatrix}
-\beta_{(j)} \\
\alpha_{(j)} \\
-\bar{\beta}_{(j)} \\ 
\bar{\alpha}_{(j)}
\end{pmatrix}  
\qquad & 
{\rm for} \, x \rightarrow + \infty, 
\end{align}
where $k_{(\pm)}$ and $\kappa_{(\pm)}$ are given by 
$\hbar k_{(\pm)} = \sqrt{ m \left [  
\sqrt{( c_{+} n \pm C )^{2}+4E^{2}} - ( c_{+} n \pm C)
\right ] }$ and $
\hbar \kappa_{(\pm)} =  \sqrt{ m \left [ 
\sqrt{( c_{+} n \pm C )^{2}+4E^{2}} + ( c_{+} n \pm C ) 
\right ] }$, respectively. 
Details of excitations for the uniform system 
[such as the wave numbers of the oscillating and evanescent waves, 
$k_{\pm}$ and $\kappa_{\pm}$, 
and the amplitudes of the wave function, 
$\alpha_{(\pm)}$, $\beta_{(\pm)}$, $\bar{\alpha}_{(\pm)}$, and $\bar{\beta}_{(\pm)}$], 
are summarized in Appendix~\ref{AppendixA}. 
The mode specified by $k_{(+)}$ $(\kappa_{(+)})$ or $k_{(-)}$ ($\kappa_{(-)}$) denotes 
the out-of-phase mode or in-phase mode between the spin components $\pm 1$. 
Its energy $E_{(\pm)}$ is given by $E_{(\pm)} = \sqrt{\varepsilon ( \varepsilon + c_{+} n \pm C ) }$, 
where $\varepsilon \equiv \hbar^{2}k_{(\pm)}^{2}/(2m)$ and $C \equiv \sqrt{(c_{-} n)^{2}  + 4c_{0}c_{1}n^{2}A^{2}}$ with $A=g\mu_{\rm B}B/(c_1 n)$. 
In Ref.~\cite{Ohmi1998}, 
these excitations are regarded as the collective modes coupled with the number density and the longitudinal spin density. 
These excitations are gapless, and the inherent energy gap, $\Delta_{B= 0} = 0$, is zero. 
These excitations in the low-energy limit are regarded as phase modes of condensates, as seen in the Bogoliubov mode, 
so that we name the mode with $E_{(+)}$ $[E_{(-)}]$, ``Bogoliubov and longitudinal spin excitation [put-of-phase (in-phase)]'' as in Table~\ref{tableI}.

The transmission coefficient of the excitation $\tilde{\phi}_{0}$ 
is, however, obtained by solving the following equations: 
\begin{align}
E 
\begin{pmatrix}
u_{0} \\ v_{0} 
\end{pmatrix}
= 
\begin{pmatrix}
\hat{\mathcal H}_{0}  & - 2 c_{1} \Phi_{+1} \Phi_{-1} \\
2 c_{1} \Phi_{+1} \Phi_{-1} &  -\hat{\mathcal H}_{0}
\end{pmatrix}
\begin{pmatrix}
u_{0} \\ v_{0} 
\end{pmatrix}, 
\label{unsatuBogoZeroS}
\end{align}
where $\hat{\mathcal H}_{0} \equiv  - {\displaystyle \frac{\hbar^{2}}{2m} \frac{d^{2}}{dx^{2}}} + V_{\rm ext} - c_{0} n + c_{+} (\Phi_{+1}^{2} + \Phi_{-1}^{2})$, 
and we used the Bogoliubov-type transformation given by $\tilde{\phi}_{0} = u_{0} e^{-iEt/\hbar} - v_{0}^{*} e^{+iEt/\hbar}$. 
We impose the following boundary conditions: 
\begin{align}
\begin{pmatrix}
u_{0} \\ v_{0} 
\end{pmatrix}
= & 
e^{ikx}
\begin{pmatrix}
\alpha_{0} \\
\beta_{0} 
\end{pmatrix} 
+ 
r
e^{-ikx}
\begin{pmatrix}
\alpha_{0} \\
\beta_{0}
\end{pmatrix}  
+ 
a
e^{\kappa x}
\begin{pmatrix}
- \beta_{0} \\
\alpha_{0}
\end{pmatrix} 
\qquad & 
{\rm for} \, x \rightarrow - \infty, 
\\
\begin{pmatrix}
u_{0} \\ v_{0} 
\end{pmatrix}
= & 
t 
e^{ik x}
\begin{pmatrix}
\alpha_{0} \\
\beta_{0}
\end{pmatrix}  
+ 
b_{0} 
e^{-\kappa x}
\begin{pmatrix}
- \beta_{0} \\
\alpha_{0} 
\end{pmatrix} 
\qquad & 
{\rm for} \, x \rightarrow + \infty, 
\end{align} 
where $k$ and $\kappa$ are given by 
$\hbar k = \sqrt{ 2m \left [ \sqrt{( c_{1}n )^{2} (1-A^{2}) + E^{2} } -c_{1}n \right ]}$, and 
$\hbar \kappa = \sqrt{ 2m \left [ \sqrt{( c_{1}n )^{2} (1-A^{2}) + E^{2} } + c_{1}n \right ]}$, respectively. 
(Details of this excitation for the uniform regime are also given in Appendix~\ref{AppendixA}.) 
This mode is associated with the transverse spin wave mode 
with the energy $E = \sqrt{\varepsilon (\varepsilon + 2c_{1} n)+  (g\mu_{\rm B}B)^{2}}$. 
This mode has the energy gap $ g\mu_{\rm B}B$, but the inherent energy gap 
is zero (i.e., $\Delta_{B = 0} = 0$).

\begin{figure}[tbp]
\begin{center}
\includegraphics[width=8.5cm]{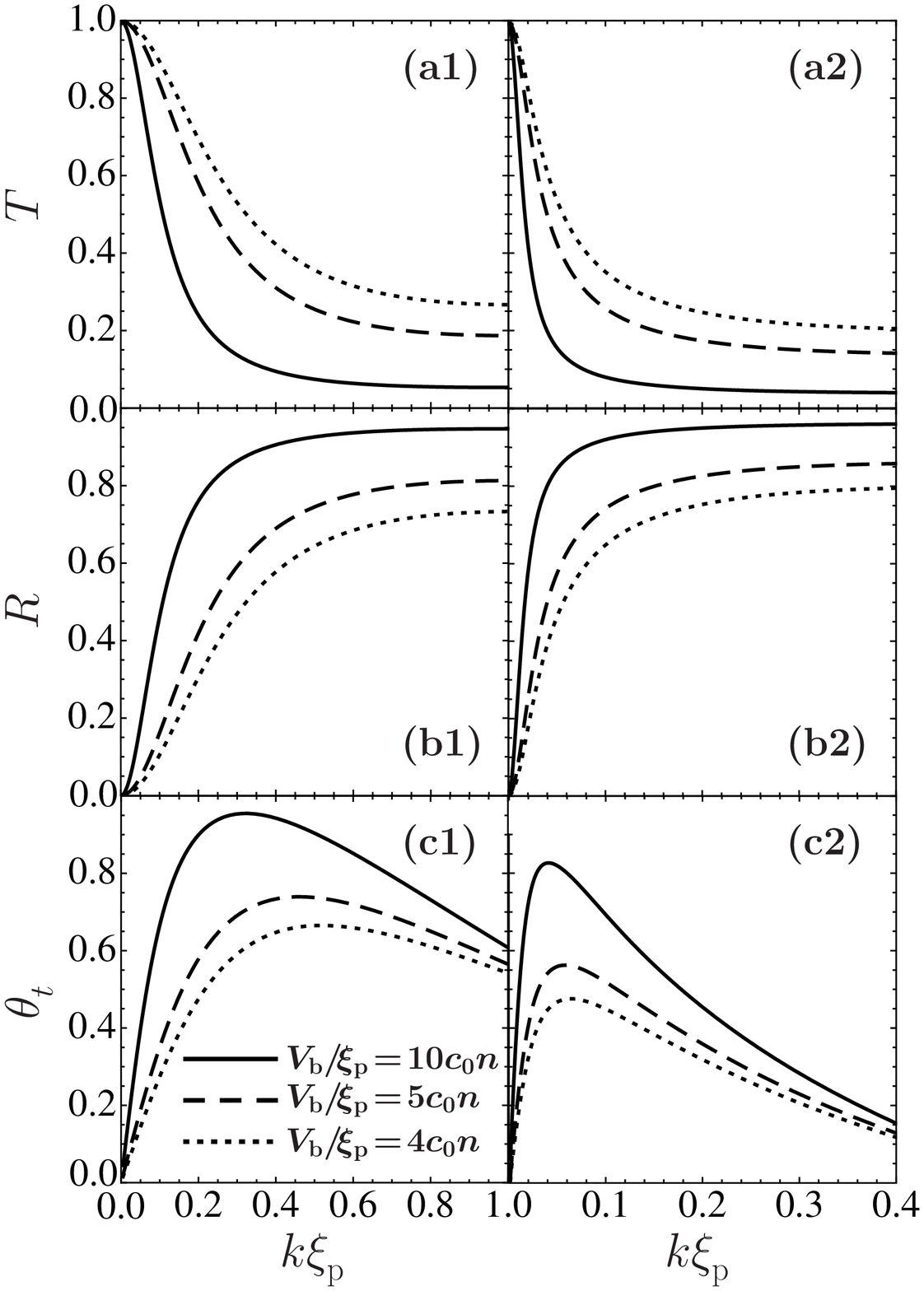}
\end{center}
\caption{
Tunneling properties of out-of-phase and in-phase modes in the unsaturated magnetization state through the barrier $V_{\rm ext} (x) = V_{\rm b}\delta (x)$. 
(a1) The transmission coefficient $T$, (b1) the reflection coefficient $R$, and (c1) the phase shift of the out-of-phase mode 
for the case where the incoming mode is the out-of-phase mode. 
The transmission and reflection coefficients of the in-phase mode are not plotted. 
We found that these values can be regarded as zero, and one can find $T + R = 1$. 
(a2) The transmission coefficient $T$, (b2) the reflection coefficient $R$, and (c2) the phase shift of the in-phase mode 
for the case where the incoming mode is the in-phase mode. 
We also do not plot the transmission and reflection coefficients of the out-of-phase mode in this case. 
We also found that these values are regarded as zero, and one can find $T + R = 1$. 
We use the parameter of the magnetic field as $g\mu_{\rm B}B = 0.5 c_{1}n$, 
and the scattering lengths $a_{0} : a_{2} = 46 : 52$, following Ref.~\cite{Ho1998}. 
}
\label{fig8}
\end{figure}

\begin{figure}[tbp]
\begin{center} 
\includegraphics[width=6cm]{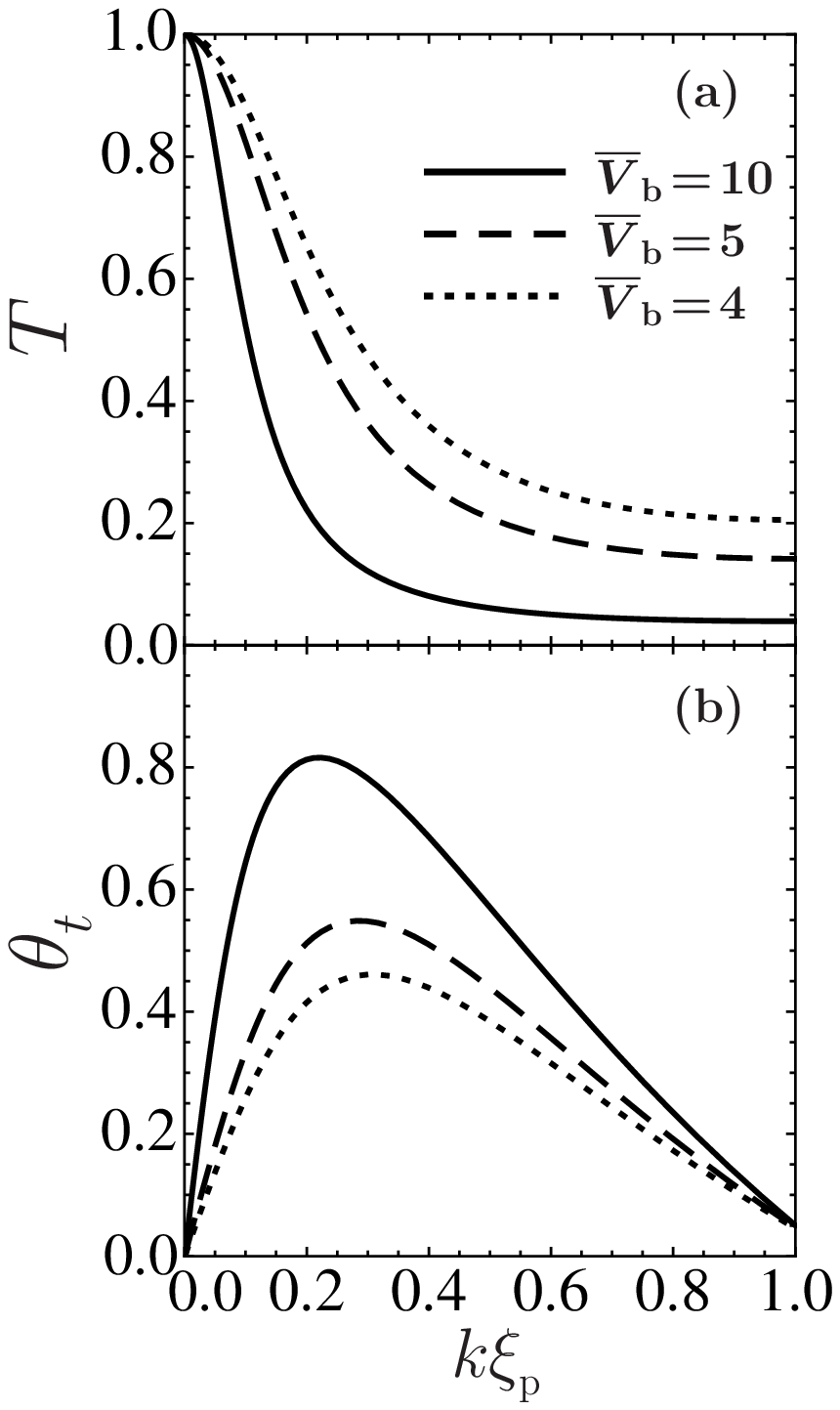}
\end{center}
\caption{
(a) Transmission coefficients and (b) corresponding phase shifts of the transverse spin-wave mode in the unsaturated magnetization state at $B\neq 0$. 
We use the same potential barrier and the same parameters as in Fig.~\ref{fig8}. 
}
\label{fig9}
\end{figure}

Figure~\ref{fig8} plots the results for the incident mode being the out-of-phase mode (left panels) 
and the in-phase mode (right panels); 
Fig.~\ref{fig9} plots the results of the transverse spin wave mode. 
Here the wave number $k$ is scaled as $k\xi_{\rm p}$, where $\xi_{\rm p} \equiv \hbar/\sqrt{mc_{0}n}$. 
We also find that all excitations tunnel through the potential barrier in the long wavelength limit, 
where phase shifts of all modes reach zero. 
We find that the sum of $T$ and $R$ for the out-of-phase (in-phase mode) is unity, as seen in Fig.~\ref{fig8}. 
As a result, when the incident mode is the out-of-phase (in-phase) mode, 
the transmitted and reflected waves are the out-of-phase (in-phase) mode itself 
and do not include the opposite type of mode [i.e., the in-phase (out-of-phase) mode]~\cite{Numerical}. 

\begin{figure}[tbp]
\begin{center}
\includegraphics[width=8.5cm]{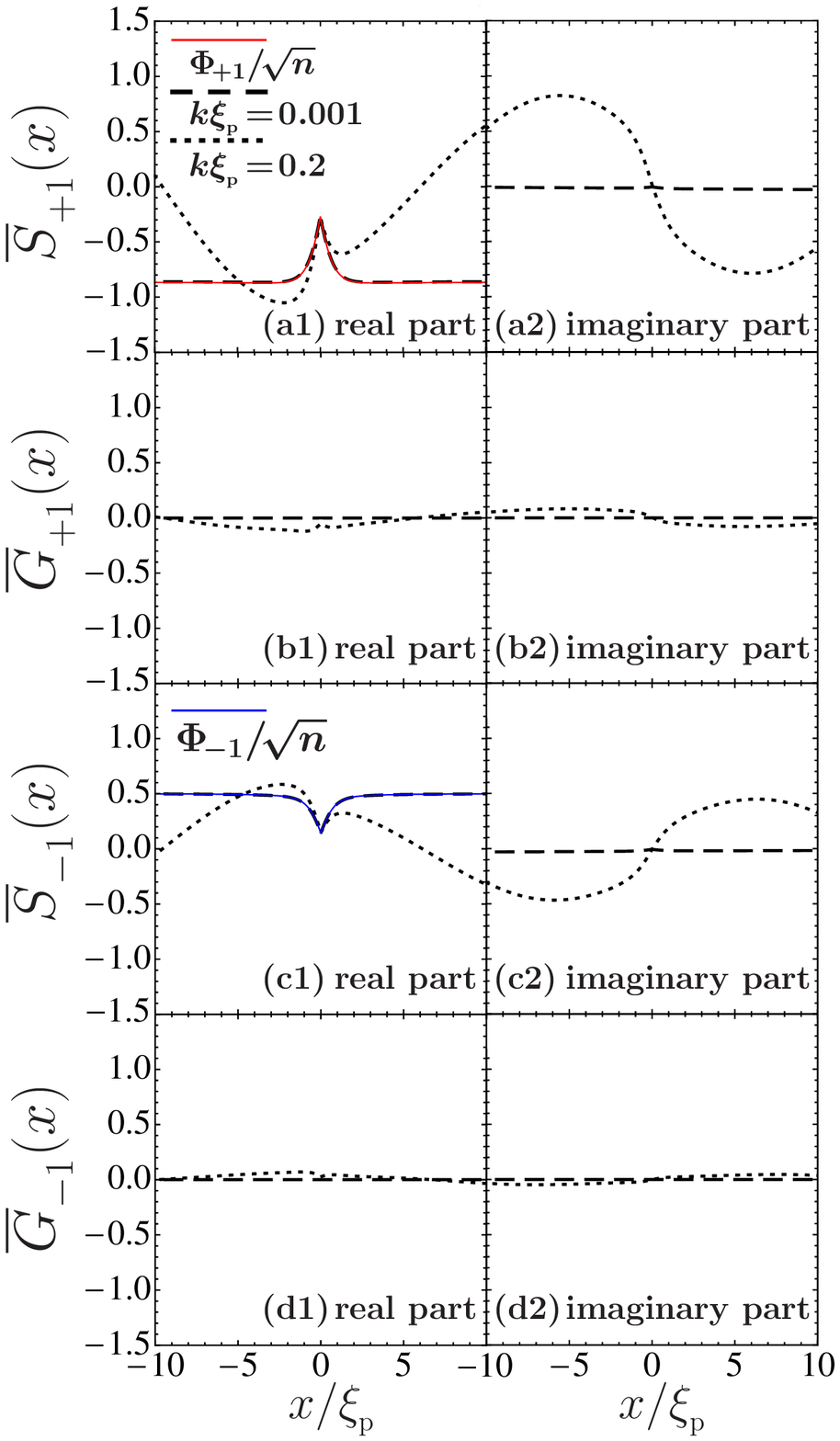}
\end{center}
\caption{(Color online) 
Wave functions when the incoming mode is the out-of-phase mode in the unsaturated magnetization state. 
Normalized functions $\overline{S}_{\pm1}$ and $\overline{G}_{\pm1}$, respectively, are given by 
$\overline{S}_{\pm1} = {\mathcal N}_{\pm1}^{(+)}{S}_{\pm1}$ 
and $\overline{G}_{\pm1} = {\mathcal N}_{\pm1}^{(+)} {G}_{\pm1}$, 
where ${\mathcal N}_{\pm 1}^{(+)} \equiv \sqrt{(1\pm A)/2}\sqrt{E C/[ (C \pm c_{+} A)(c_{+} + C) ]}$. 
We use the $\delta$-function potential barrier $V_{\rm ext} (x) = V_{\rm b} \delta (x)$ with $V_{\rm b}/\xi_{\rm p} = 3 c_{0}n$, 
and the magnetic field given by $g\mu_{\rm B}B = 0.5 c_{1}n$. 
Parameters $a_{0} : a_{2} = 46 : 52$ are used, following Ref.~\cite{Ho1998}.  
}
\label{fig10}
\end{figure}

\begin{figure}[tbp]
\begin{center}
\includegraphics[width=8.5cm]{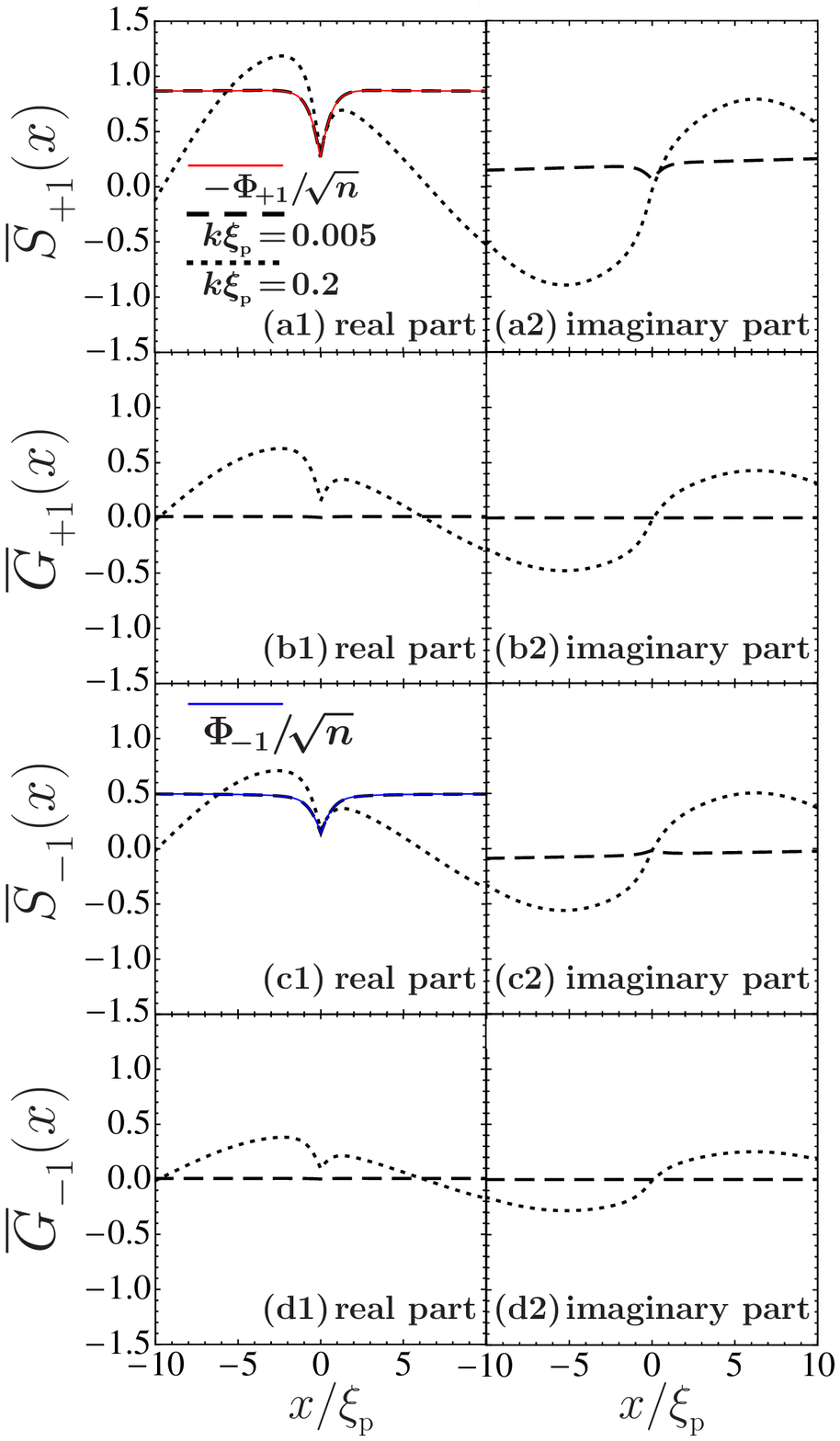}
\end{center}
\caption{(Color online) 
Wave functions when the incoming mode is the in-phase mode in the unsaturated magnetization state. 
Normalized functions $\overline{S}_{\pm1}$ and $\overline{G}_{+1}$, respectively, given by 
$\overline{S}_{\pm1} = {\mathcal N}_{\pm1}^{(-)}{S}_{\pm1}$ 
and $\overline{G}_{\pm1} = {\mathcal N}_{\pm1}^{(-)} {G}_{\pm1}$, 
where ${\mathcal N}_{\pm1}^{(-)} \equiv \sqrt{(1\pm A)/2}\sqrt{E C/[C \mp c_{+} A)(c_{+}-C)]}$. 
We use the same parameters and potential barrier as in Fig.~\ref{fig10}. 
}
\label{fig11}
\end{figure}

As in Sec.~\ref{SecIII}, 
we saw that the wave function of an excitation has the same form as the condensate wave function when the anomalous tunneling phenomenon occurs 
in the saturated magnetization state. 
Here, let us examine the low-momentum properties of the excitations in the unsaturated magnetization state. 
Figures~\ref{fig10} and \ref{fig11}, respectively, show wave functions of the out-of-phase mode and the in-phase mode. 
As for the out-of-phase mode, one finds that $S_{+1}$ and $S_{-1}$ are, respectively, proportional to $\Phi_{+1}$ and $\Phi_{-1}$; 
but $G_{\pm 1}$ is absent in the low-momentum limit. 
Consequently, from Fig.~\ref{fig10}, the solution in the long-wavelength limit for the out-of-phase mode is given 
by $(u_{+1},v_{+1},u_{-1}, v_{-1}) \propto ( \Phi_{+1}, \Phi_{+1},\Phi_{-1},\Phi_{-1})$.  
Because of the opposite sign between $\Phi_{\pm 1}$ as in Fig.~\ref{fig7}, this excitation can be confirmed to be the out-of-phase mode. 

As for the in-phase mode, one first finds that $G_{\pm 1}$ is absent for the small momentum from Fig.~\ref{fig11}. 
Also from Fig.~\ref{fig11}, however, one finds that ${\rm Im}[S_{\pm 1}]$ is not absent, 
although ${\rm Re}[{S}_{\pm 1}]$ is proportional to $\Phi_{\pm1}$. 
This is due to the momentum $k\xi_{\rm f} = 0.001$ being not as small as the low momentum limit for the in-phase mode. 
In fact, ${\rm Im}[S_{\pm 1}]$ has a gradient in the asymptotic regime. 
In the low momentum limit, the gradient of the wave function for the tunneling problem should be absent far from the potential barrier, 
because the boundary condition of this problem is given by $\exp{(ikx)} + r \exp{(-ikx)}$ for $x \ll -\xi_{\rm p}$, 
and $t \exp{(ikx)}$ for $x \gg \xi_{\rm p}$, 
so that this condition leads to $1+r (= {\rm const.})$ for $x \ll -\xi_{\rm p}$ and $t (= {\rm const.})$ for $x \gg \xi_{\rm p}$ in the low-momentum limit. 
To examine the wave function $S_{\pm 1}$ of this mode in the low-momentum limit, we return to Eqs. (\ref{EqSGPM48}) and (\ref{EqSGPM50}). 
In the limit $E\rightarrow 0$ (i.e., $k\rightarrow 0$), 
Eqs. (\ref{EqSGPM48}) and (\ref{EqSGPM50}) lead to 
\begin{align}
0 = & \hat{\mathcal H}_{\pm} S_{\pm 1}. 
\label{ZeroLimitPolarSpm1} 
\end{align} 
If we compare (\ref{GPPolarFM+1}) with (\ref{ZeroLimitPolarSpm1}), 
a set of linearly independent solutions to (\ref{ZeroLimitPolarSpm1}) is found to be given by 
$\Phi_{\pm 1}(x)$ and $\Phi_{\pm 1} (x) \int_{0}^{x} dx' \Phi_{\pm 1}^{-2}(x')$. 
The solution $S_{\pm 1}$ is now given by the linear combination of these two functions. 
However, since $\Phi_{\pm1} = \mp\sqrt{1\pm A} (= {\rm const.})$ for $|x| \gg \xi_{\rm p}$, 
the second solution is proportional to $x$ for large $|x|$. 
This means that the second solution does not satisfy the boundary condition of the tunneling problem in the low-momentum limit, and is found to be absent. 
As a result, $S_{\pm 1} \propto \Phi_{\pm 1}$ follows.  
Consequently, the solution in the long-wavelength limit is given by $(u_{+1},v_{+1},u_{-1}, v_{-1}) \propto ( - \Phi_{+1}, - \Phi_{+1},\Phi_{-1},\Phi_{-1})$, 
where the sign is given by the boundary condition.

\begin{figure}[tbp]
\begin{center}
\includegraphics[width=8.5cm]{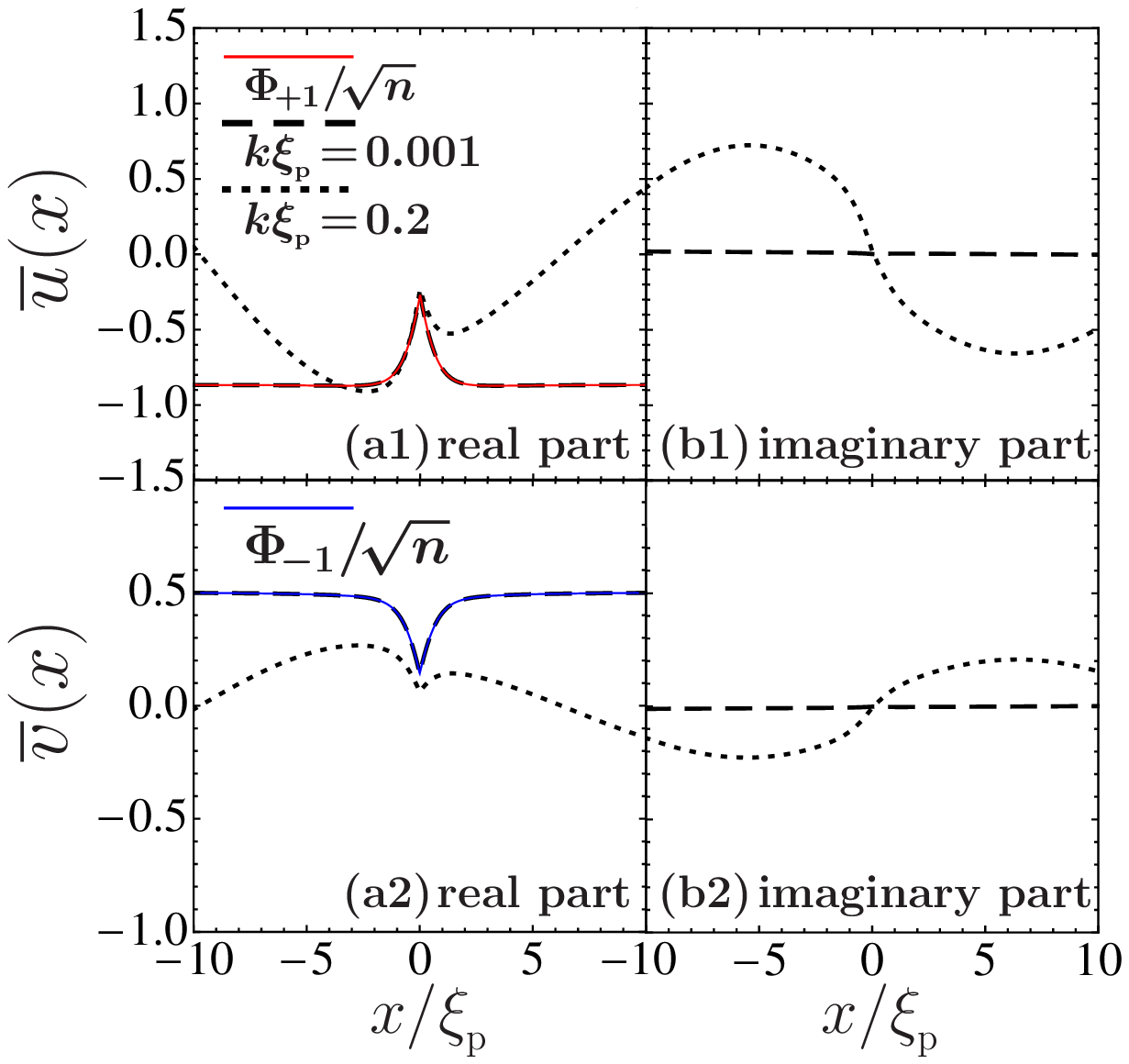}
\end{center}
\caption{
(Color online) 
Wave functions of the spin component $0$ in the unsaturated magnetization state. 
Normalized functions $\overline{u}$ and $\overline{v}$ are 
given by $(\overline{u}, \overline{v}) \equiv  -\sqrt{A/n}(u_{0}, v_{0})$. 
The potential barrier and parameters are the same as those in Fig.~\ref{fig10}. 
}
\label{fig12}
\end{figure}

We finally discuss the transverse spin excitation. 
Figure~\ref{fig12} plots the spatial variation of wave functions $u_{0}$ and $v_{0}$. 
From this figure, we find that the wave function of the excitation in the low-momentum limit has the same form as the condensate wave function 
[i.e., $-\sqrt{A}({u}_{0}, {v}_{0}) = (\Phi_{+1}, \Phi_{-1})$]. 
This property can be confirmed from (\ref{unsatuBogoZeroS}) 
in the limit $k\rightarrow 0$ (i.e., $E\rightarrow g\mu_{\rm B} B$). 
In this limit, one finds that $u_{0}$ and $v_{0}$ have solutions $u_{0} \propto \Phi_{+1}$ 
and $v_{0} \propto \Phi_{-1}$ by 
comparing (\ref{unsatuBogoZeroS}) with (\ref{GPPolarFM+1}). 
From those results, $(u_{0}, v_{0})  \propto (\Phi_{+1}, \Phi_{-1})$ follows in the low momentum limit. 

In summary, recalling the case of the saturated magnetization state, 
we found that all excitations showing total transmission in the long-wavelength limit 
have wave functions equal to the condensate wave functions. 
As for the Bogoliubov mode in the scalar BEC, 
Ref.~\cite{Fetter1972} pointed out the existence of the solution equal to the condensate wave function for the zero mode. 
The zero modes in the spin-1 BEC also have the same property; 
their wave functions in the low momentum limit are equal to the condensate wave functions for the spin-1 BEC. 

\section{Excitations in the saturated magnetization state in the presence of the $\delta$-function potential barrier}\label{SecV}

In this section, we examine solutions in the presence of the $\delta$-function potential barrier, 
focusing on spin modes in the saturated magnetization state. 
As for the Bogoliubov excitation, the property in the current free state was studied in Ref.~\cite{Kovrizhin2001}. 

First, we derive an analytic form of the transmission coefficient for $\tilde{\phi}_{0}$. 
We assume the potential energy $V_{\rm ext}(x)$ as $V_{\rm L}$ for $x < 0$, 
$V_{\rm R}$ for $x > 0$, and $V_{\rm b}\delta(x)$ for $x=0$. 
As shown in Eq. (\ref{satuGPP}), 
the GP-type equation in this saturated magnetization state is given by 
\begin{align}
\left [ - \frac{\hbar^{2}}{2m} \frac{d^{2}}{dx^{2}} + V_{\rm ext}(x) - \mu
+ c_{+} {\Phi}_{+1}^{2}  \right ] {\Phi}_{+1} = 0. 
\end{align} 
Let us introduce the healing length $\xi_{\rm L (R)}$, 
given by $\xi_{\rm L (R)} \equiv \sqrt{m(\mu - V_{\rm L (R)})}/\hbar$. 
In the asymptotic regime, we have the following forms: 
$\Phi_{+1} = \sqrt{(\mu -V_{\rm L})/c_{+}} (\equiv \sqrt{n_{\rm L}})$ 
for $x \ll - \xi_{\rm L}$, 
and 
$\Phi_{+1} =\sqrt{(\mu -V_{\rm R})/c_{+}} (\equiv \sqrt{n_{\rm R}})$ 
for $x \gg \xi_{\rm R}$. 
In the presence of the $\delta$-function potential barrier, 
the condensate wave function is given by 
\begin{eqnarray}
\Phi_{+1} 
= 
\left \{
\begin{array}{ll}
\displaystyle{
\sqrt{n_{\rm L}}
} 
\tanh{[(-x+x_{0,{\rm L}})/\xi_{\rm L}]} \quad & {\rm for} \quad x<0, 
\\
\sqrt{n_{\rm R}} \tanh{[(x+x_{0,{\rm R}})/\xi_{\rm R}]} \quad & {\rm for} \quad x \geq 0. 
\end{array} 
\right .
\end{eqnarray}
Parameters $x_{0, \rm L}$ and $x_{0, \rm R}$ are determined by 
sewing the wave functions at the barrier 
[i.e., 
$\Phi_{+1}(+0) = \Phi_{+1}(-0)$ and 
$\Phi_{+1}' (+0) - \Phi_{+1}' (-0) = 2mV_{\rm b}\Phi_{+1} (0) /\hbar^{2}$
].

In this case, the solution to (\ref{Spin1ExcitationFerroEq0}) is given by 
\begin{align}
\tilde{\phi}_{0}(x) =& f_{\rm L}(k_{}, x) \exp{(+ ik_{} x)}
+ r f_{\rm L}(-k_{}, x) \exp{(-ik_{} x)} \qquad & {\rm for} \quad x < 0, \\
\tilde{\phi}_{0}(x) =& t f_{\rm R}(k_{}, x) \exp{(+ ik_{}x)} \qquad &{\rm for} \quad x \geq 0,
\end{align}
where $f_{\nu}(k, x) \equiv -ik\xi_{\nu}{\rm sgn}(x) 
+ \tanh{ [( |x|+  x_{0, \nu})/\xi_{\nu} ] }$. 
Sewing the different pieces of the wave functions together at $x=0$
[i.e., $\tilde{\phi}_{0}(+0) = \tilde{\phi}_{0}(-0)$ and 
$\tilde{\phi}_{0}' (+0) - \tilde{\phi}_{0}' (-0) = 2mV_{\rm b}\tilde{\phi}_{0} (0)/\hbar^{2}$], 
we have the amplitude transmission and reflection coefficients $t$ and $r$, given by 
\begin{align}
t = &\frac{2\eta_{\rm L}\xi_{\rm L}\xi_{\rm R}(1+k_{}^{2}\xi_{\rm L}^{2})}
{(\eta_{\rm L} - ik_{}\xi_{\rm L})(\xi_{\rm L}^{2} + \xi_{\rm R}^{2} 
- 2ik_{}\eta_{\rm L}\xi_{\rm L}\xi_{\rm R}^{2})}, 
\label{eq:deltaspint}
\\
r = &\frac{
2i k_{} \eta_{\rm L}^{2}\xi_{\rm L}\xi_{\rm R}^{2} 
-ik_{} (\xi_{\rm L}^{2} + \xi_{\rm R}^{2}) -\eta_{\rm L}(\xi_{\rm L}^{2} - \xi_{\rm R}^{2})
}
{(\eta_{\rm L} - ik_{}\xi_{\rm L})(\xi_{\rm L}^{2} + \xi_{\rm R}^{2} 
- 2ik_{}\eta_{\rm L}\xi_{\rm L}\xi_{\rm R}^{2})}, 
\label{eq:deltaspinr}
\end{align}
where $\eta_{\rm L} \equiv \tanh{(x_{0, \rm L}/\xi_{\rm L})}$. 
In the long-wavelength limit, 
we have the transmission coefficient $T = |t|^{2}$ in terms of the magnetization $M_{\nu} \equiv \mu_{\rm B}n_{\nu}$, 
which is given by $T = 4M_{\rm L}M_{\rm R}/(M_{\rm L} + M_{\rm R})^{2}$. 
This result recovers the result in Ref.~\cite{WatabeKatoLett}. 
We conclude that the transverse spin-wave mode in the saturated magnetization state 
shows partial transmission in the long-wavelength limit when $M_{\rm L} \neq M_{\rm R}$. 

In the case where $M_{\rm L} = M_{\rm R}$, 
we have 
\begin{align}
t = \frac{\eta (1+k_{}^{2}\xi^{2})}{(\eta - ik_{}\xi)(1- i k_{}\eta\xi)}, 
\\
r = 
\frac{ik_{}(\xi \eta^{2} - 1 )}{(\eta - ik_{}\xi)(1- i k_{}\eta\xi)}, 
\end{align}
where $\eta \equiv \eta_{\rm L}$ and $\xi \equiv \xi_{\rm L } = \xi_{\rm R}$. 
As a result, total transmission in the limit $k_{} \rightarrow 0$ occurs. 
In the high-barrier limit, we have $\eta \simeq \hbar^{2}/(m\xi V_{\rm b})$, 
so that the amplitude transmission coefficient in the long-wavelength regime is given by $t \simeq 1 + ik_{} m\xi^{2} V_{\rm b}/\hbar^{2}$. 
In this limit, the transmission coefficient $T$ is given by $T = |t|^{2} = 1 + {\mathcal O}(k_{}^{2})$. 
We also find $\tilde{\phi}_{0}(x) = \Phi_{+1} (x)$ up to an overall normalization factor in the limit $k_{}\rightarrow 0$. 
The correspondence between wave functions holds, as discussed in Sec.~\ref{SecIII}. 

We note that it is easy to extend this one-dimensional tunneling problem to the problem of the reflection and the refraction of the excitation as studied in~\cite{Watabe2008}. 
[The barrier has only the $x$ dependence (i.e., the condensate wave function also has only the $x$ dependence], 
and the incoming mode has the incident angle defined by that between the $x$-direction and the incident momentum.) 
Even if $V_{\rm L} \neq V_{\rm R}$, 
the energy spectrum $E = \hbar^{2}k^{2}/(2m) + g\mu_{\rm B}B$ is independent of the potential barrier. 
Because of the translational invariance, the momenta along the wall on both sides are equal. 
As a result, the equation of $\tilde{\phi}_{0}$ is obtained when we replace $E$ in Eq.~(\ref{Spin1ExcitationFerroEq0}) 
with $\hbar^{2}k_{x}^{2}/(2m) + g\mu_{\rm B}B$, where $\hbar k_{x}$ is the momentum along the $x$-axis. 
Consequently, one finds no refraction, and the $k_{x}$-dependence of the amplitude transmission and reflection coefficients is obtained if one replaces $k$ 
in (\ref{eq:deltaspint}) and (\ref{eq:deltaspinr}) with $k_{x}$. 

The $\delta$-function potential barrier is also useful 
to study the total reflection of the quadrupolar spin mode in the saturated magnetization state. 
In what follows, we investigate a simple model: 
the junction of BECs with equal densities separated by 
the potential barrier $V_{\rm ext} (x) = V_{\rm b}\delta(x)$. 
We compare the properties of wave functions $\tilde{\phi}_{0}$ with  $\tilde{\phi}_{-1}$ in the presence of this potential barrier. 
The simple problem with the $\delta$-function potential barrier has revealed essential phenomena for the tunneling problem of the Bogoliubov excitation~\cite{Kovrizhin2001, Danshita2006}. 
The implication of the tunneling problem against the $\delta$-function potential barrier is expected to be common in problems for any potential barriers with an arbitrary shape. 

The GP-type equation (\ref{satuGPP}) normalized by $ c_{+} n$ is given by 
\begin{align}
\left [ - \frac{1}{2} \frac{d^{2}}{dx^{2}} + V_{\rm b}\delta(x) + \phi^{2} -1 \right ] \phi = 0, 
\end{align}
where we replaced $x/\xi_{\rm f}$ with $x$, and $V_{\rm b}/(\xi_{\rm f} c_{+} n)$ with $V_{\rm b}$. 
In this case, the solution is given by 
$\phi (x) = \tanh(|x| + x_{0})$, 
where the parameter $x_{0}$ is determined by $\eta \equiv \tanh(x_{0}) = - (V_{\rm b}/2) + \sqrt{(V_{\rm b}/2)^{2} + 1}$ through the boundary condition at $x=0$ [i.e., $\phi(+0) = \phi(-0) \equiv \eta$ and $\phi'(+0) = \phi'(-0) + 2V_{\rm b}\phi(0)$]. 
From Eqs. (\ref{Spin1ExcitationFerroEq0}) and (\ref{Spin1ExcitationFerroEqMinus}), 
an equation of $\tilde{\phi}_{j}$ for $j = 0$ or $-1$ at $|x|>0$ can be reduced to  
\begin{align}
\left [ - \frac{1}{2} \frac{d^{2}}{dx^{2}} +  A_{j}(\phi^{2} -1) \right ] \tilde{\phi}_{j} = 
\frac{k_{}^{2}}{2} \tilde{\phi}_{j}, 
\label{satuGPNormalize}
\end{align}
where $A_{0} = 1$ and $A_{-1} = 1-2\tilde{c}_{1}$ with $\tilde{c}_{1}$ being $c_{1}/c_{+}$. 

Using the fact $d\phi(x)/dx = {\rm sgn}(x)[1-\phi^{2}(x)]$, 
we transform Eq. (\ref{satuGPNormalize}) into the following form: 
\begin{align}
(1 - \phi^{2})\frac{d^{2} \tilde{\phi}_{j}}{d \phi^{2}} - 2 \phi \frac{d\tilde{\phi}_{j}}{d\phi} 
+ \left ( 2 A_{j} + \frac{k_{}^{2}}{1-\phi^{2}} \right ) \tilde{\phi}_{j} = 0. 
\label{Eq74:2010/11/5}
\end{align}
A general solution to (\ref{Eq74:2010/11/5}) is given by 
\begin{align}
\tilde{\phi}_{j} = \alpha_{j}^{\nu} (k_{}) P_{\zeta_{j}}^{ik_{}}(\phi) 
+ \beta_{j}^{\nu} (k_{}) Q_{_{\zeta_{j}}}^{ik_{}}(\phi), 
\end{align}
where $P_{\zeta_{j}}^{ik_{}}$ and $Q_{\zeta_{j}}^{ik_{}}$ are associated Legendre functions, 
and a parameter $\zeta_{j}$ is given by $\zeta_{j} \equiv (-1+\sqrt{1+8A_{j}})/2$ 
[i.e., $\zeta_{0} = 1$ and $\zeta_{-1} = (-1+\sqrt{9-16\tilde{c}_{1}})/2$]. 
The index $\nu$ in coefficients $\alpha_{j}^{\nu}$ and $\beta_{j}^{\nu}$ represents ${\rm L}$ for $x \leq 0$ and ${\rm R}$ for $x \geq 0$. 

An associated Legendre 
function $Q_{\zeta_{j}}^{ik_{}\rightarrow 0} (\phi)$ diverges at $|x| \rightarrow \infty$ 
because of $\phi(|x| \rightarrow \infty) = 1$, 
and hence we have a condition $\beta_{j}^{\nu}(k_{}\rightarrow 0) = 0$. 
Sewing the different pieces of the wave functions together at $x=0$ 
[i.e., 
$\tilde{\phi}_{j} (x = +0) = \tilde{\phi}_{j} (x = -0)$ and 
$ {\tilde{\phi}_{j}}' (+0) -  {\tilde{\phi}_{j}'} (-0) 
= 
2V_{\rm b} \tilde{\phi}_{j} (0)$], 
we have two conditions: $\alpha_{j} \equiv \alpha_{j}^{\rm L} (k_{}\rightarrow 0) = \alpha_{j}^{\rm R} (k_{}\rightarrow 0) $, 
and 
\begin{align}
0 = \alpha_{j} \lim\limits_{\phi \rightarrow \eta} \lim\limits_{k_{} \rightarrow 0} 
\left ( \frac{dP_{\zeta_{j}}^{ik_{}} (\phi) }{d\phi}
- 
\frac{P_{\zeta_{j}}^{ik_{}} (\phi)}{\phi}
\right ). 
\label{Eq78}
\end{align}
Here we used the boundary condition of the condensate wave function at $x=0$. 

As for the mode $j=0$, $P_{\zeta_{0} (=1)}^{ik_{}\rightarrow 0}(\phi) = \phi$ holds. 
As a result, one finds that $\alpha_{0}$ is not generally zero from Eq. (\ref{Eq78}), so that $\tilde{\phi}_{0} \propto \phi$ follows. 
We here consider the boundary condition of the tunneling problem 
[i.e., $\exp{(ik_{}x)} + r(k_{})\exp{(-ik_{}x)}$ for $x\ll - 1$ and $t(k_{})\exp{(ik_{}x)}$ for $x\gg 1$]. 
In the long wavelength limit $k_{}\rightarrow 0$, 
we have $1+ r (k_{} \rightarrow 0) = t (k_{} \rightarrow 0) = \phi (|x|\rightarrow \infty) = 1$, where we set $\alpha_{0} = 1$. 
From this condition, we have total transmission $t = 1$ and $r =0$. 
We also obtain the phase shift of the amplitude transmission coefficient, 
$\theta_{t} \equiv - i \ln{(t/|t|)} = 0$.  

In the case of the quadrupolar spin mode for $\tilde{\phi}_{-1}$, $P_{\zeta_{-1} (\neq 1)}^{ik_{}\rightarrow 0}(\phi) \neq \phi$ holds. 
We conclude that $\alpha_{-1}$ should always be zero in order to satisfy the condition (\ref{Eq78}). 
As a result, 
in the long wavelength limit $k_{}\rightarrow 0$, 
we have a result $1+ r (k_{} \rightarrow 0) = t (k_{} \rightarrow 0) = 0$. 
From this condition, we have total reflection $t = 0$ and $r =-1$ in the limit $k_{} \rightarrow 0$. 
We also obtain the phase shift of the amplitude reflection coefficient 
$\theta_{r} \equiv - i \ln{(r/|r|)} = \pi$.

Modes of $\tilde{\phi}_{0}$ and $\tilde{\phi}_{-1}$ 
obey the Schr\"odinger-type equation. 
Generally, a particle obeying the Schr\"odinger equation 
shows total reflection against the potential barrier in the long-wavelength limit. 
The result of the excitation $\tilde{\phi}_{-1}$ is consistent with this well-known result. 
The potential barrier makes the amplitude of the wave function vanish at zero momentum. 
However, the excitation $\tilde{\phi}_{0}$ has a wave function with the same form as the condensate wave function, 
and its finite amplitude in the long-wavelength limit results in the finite transmission coefficient. 
This property is strongly related to the fact that the transverse spin wave is 
a Nambu-Goldstone mode. 
The barrier coupled only to the density does not lift the degeneracy of the ground state and it results in perfect transmission in the symmetric case~\cite{WatabeKatoOhashi}. 
From another point of view, 
we comment that the wave function $\tilde{\phi}_{0}$ behaves like an interacting Bose gas, 
where the potential effect disappears far from the potential barrier~\cite{Hohenberg1965}. 
However, the wave function $\tilde{\phi}_{-1}$ behaves like a free Bose gas, 
where the potential effect still remains far from it~\cite{Hohenberg1965}. 
These different properties lead to different results.

\section{Dependence on $c_{1}$ and The Linear Zeeman Effect}\label{SecVI}

\subsection{Quadrupolar spin mode in the saturated magnetization state}\label{Spin1SecQSMFS}

As seen in Fig.~\ref{fig4}, total reflection occurs in the long-wavelength limit in the quadrupolar spin mode 
in the saturated magnetization state; 
however, one finds total transmission at a small but finite momentum. 
We numerically investigate this high-transmission coefficient by changing the parameter $c_{1}$. 

\begin{figure}[tbp]
\begin{center}
\includegraphics[width=6cm]{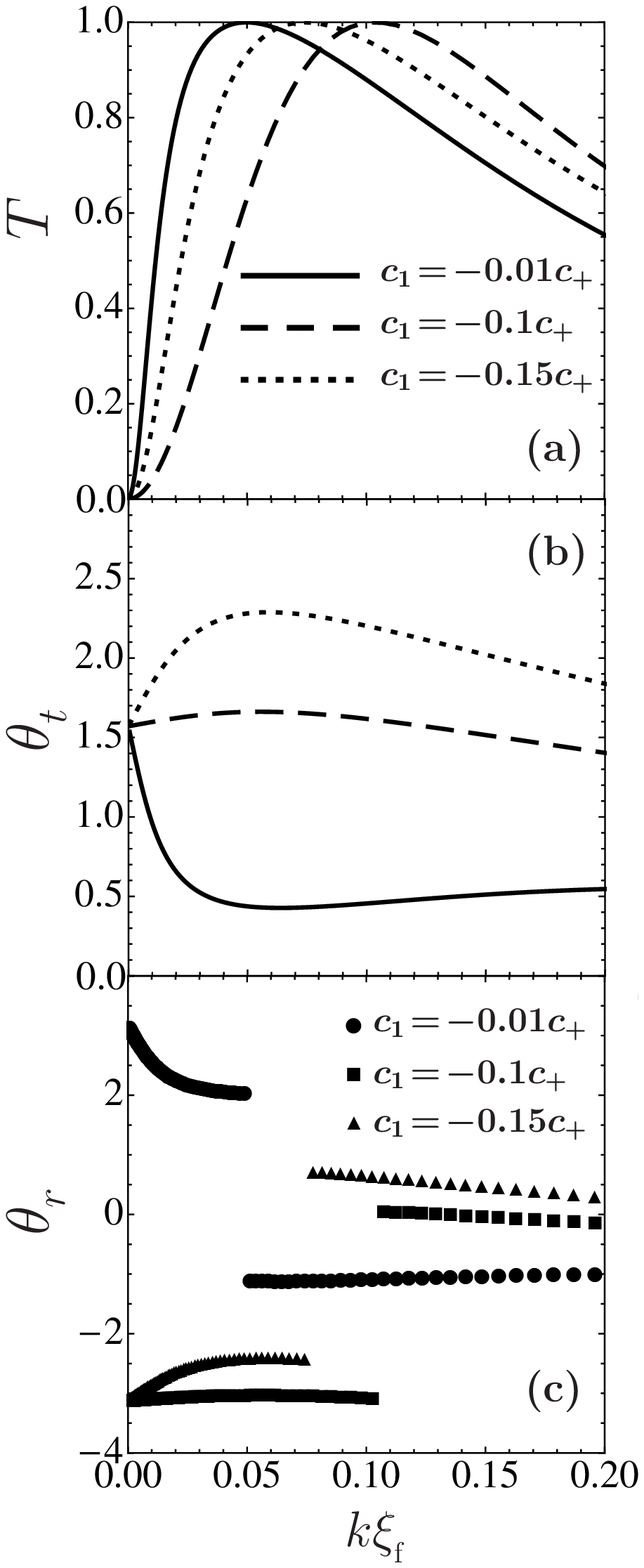}
\end{center}
\caption{
(a) Transmission coefficients, (b) corresponding phase shifts and (c) arguments of the amplitude reflection coefficients 
of the quadrupolar spin mode in the saturated magnetization state for the small coupling constant $|c_{1}|$. 
For such a small parameter, the perfect tunneling occurs at a certain momentum, 
where the argument of the amplitude reflection coefficient jumps by $\pi$. 
We call this type of property type I in this paper. 
We use the $\delta$-function potential barrier $V_{\rm ext} (x) = V_{\rm b}\delta (x)$ with $V_{\rm b}/\xi_{\rm f} = 5 c_{+} n$. 
}
\label{fig13}
\end{figure}

\begin{figure}[tbp]
\begin{center}
\includegraphics[width=6cm]{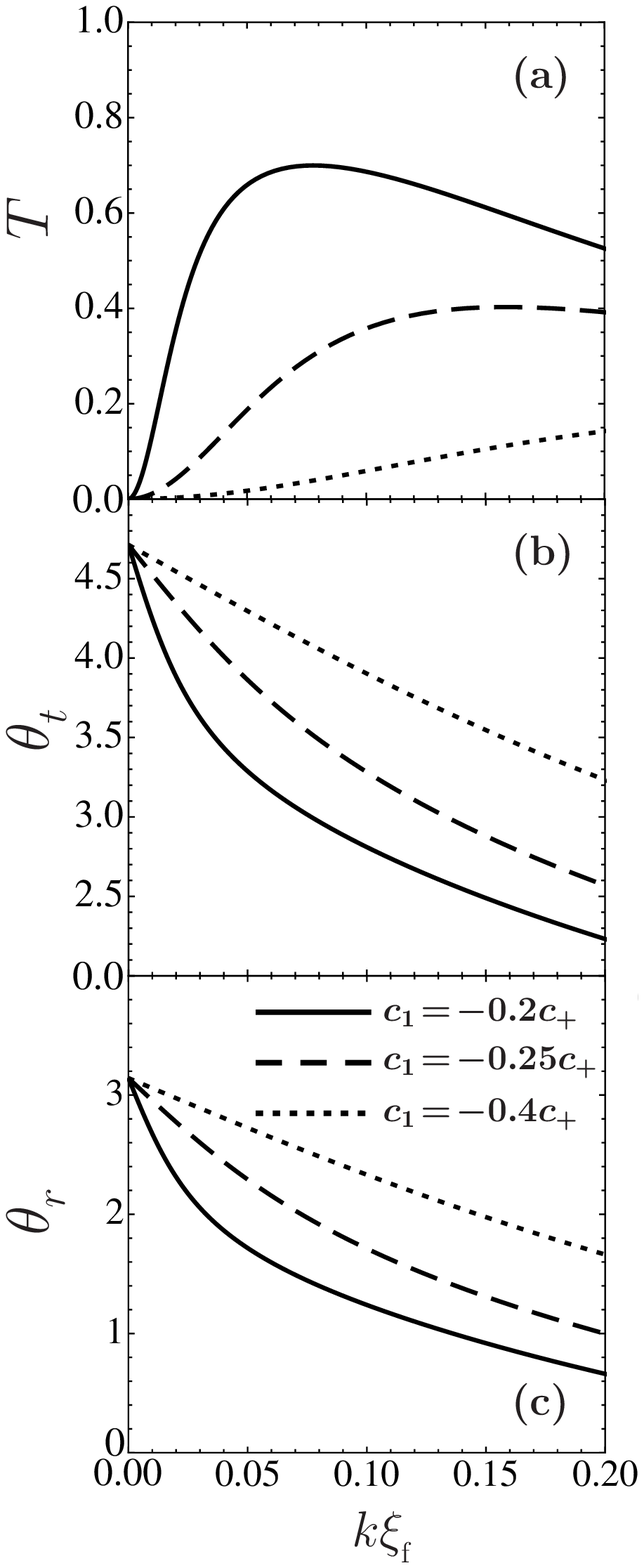}
\end{center}
\caption{
(a) Transmission coefficients, (b) corresponding phase shifts and (c) arguments of the amplitude reflection coefficients 
of the quadrupolar spin mode in the saturated magnetization state for the larger coupling constant $|c_{1}|$ than in Fig.~\ref{fig13}. 
In contrast to type I, perfect tunneling does not occur for a small momentum. 
We call this type of property type II in this paper. 
We used the same potential barrier as in Fig.~\ref{fig13}. 
}
\label{fig14}
\end{figure}

We find that the tunneling properties of the quadrupolar spin mode 
can be categorized into two types. 
The first one is the excitation whose transmission coefficient touches unity (Fig.~\ref{fig13}), which we name type I.
As seen in Fig.~\ref{fig13} (c), 
the phase of the amplitude reflection coefficient $r$ jumps by $\pi$ at a certain momentum, which corresponds to the peak position of the transmission coefficients. 
This jump is a signature of type I, since the amplitude reflection coefficient crosses zero and changes its sign. 
When we focus on the momentum reaching $T=1$ as a function of $c_{1}/c_{+}$, 
the maximum $k_{}$ is found to exist from Fig.~\ref{fig13} (a). 
The momenta reaching $T=1$ in $c_{1}/c_{+} = -0.01$ and $-0.15$ 
are less than that in the intermediate parameter $c_{1}/c_{+} = -0.1$. 
Type I can be seen in the small coupling constant regime of $c_{1}$. 
The second type is 
the excitation whose transmission coefficients do not touch unity (Fig.~\ref{fig14}), which we name type II. 
As $|c_{1}|$ increases, 
the peak value of the transmission coefficient decreases from unity. 

In type I, the phase shifts of the amplitude transmission coefficient are $\pi/2$ 
in the long wavelength limit [Fig.~\ref{fig13} (b)]. 
In type II, however, the phase shifts of the amplitude transmission coefficient are $3\pi/2$ 
in the long wavelength limit [Fig.~\ref{fig14} (b)]. 
A characteristic feature of the total reflection 
can be seen in the argument of the amplitude reflection coefficient $r$. 
The argument of the amplitude reflection coefficient $r$ 
reaches $\pm \pi$ in the long wavelength limit. 
Note that we can find type I and type II in the tunneling problem of a quantum particle obeying the Schr\"odinger equation, 
when we consider a mimic system of the present case (i.e., a system in the presence of attractive-repulsive-attractive potential)~\cite{Mimic}. 

\subsection{Linear Zeeman effect}\label{Spin1SecMC}

In the saturated magnetization state, 
the uniform magnetic field associated with the linear Zeeman effect does not change a profile of the condensate wave function. 
As for excitations, 
it changes energy intervals between eigenmodes; 
however, it does not affect wave functions of excitations when one regards them as functions of the momentum and the position. 
As a result, the tunneling properties of excitations in the saturated magnetization state are not changed by the uniform magnetic field. 

\begin{figure}[tbp]
\begin{center}
\includegraphics[width=6cm]{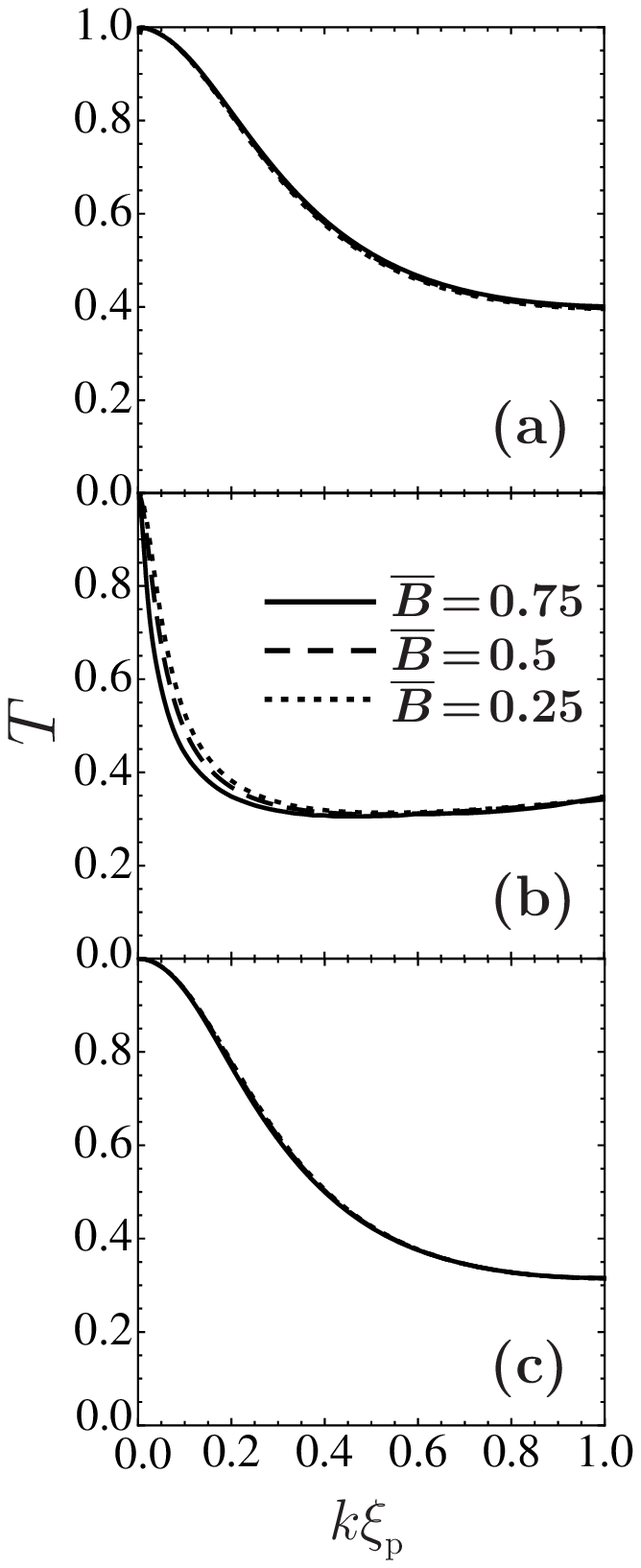}
\end{center}
\caption{
Magnetic field dependence of the transmission coefficient in the unsaturated magnetization state 
through the $\delta$-function potential barrier $V_{\rm ext}(x) = V_{\rm b}\delta (x)$ 
where $V_{\rm b}/\xi_{\rm p} = 3 c_{0}n$. 
(a), (b), and (c), respectively, show results of the out-of-phase mode, the in-phase mode, and the transverse spin mode. 
We use parameters $a_{0} : a_{2} = 46 : 52$, following Ref.~\cite{Ho1998}. $\overline{B}$ is given in Fig.~\ref{fig7}. 
}
\label{fig15}
\end{figure}

In the unsaturated magnetization state, 
the density of each component is affected 
by the magnetic field associated with the linear Zeeman effect, as shown in Fig.~\ref{fig7}. 
We thus have an issue regarding how the transmission coefficients are affected by the uniform magnetic field. 
From Fig.~\ref{fig15}, 
one finds that the effects of the uniform magnetic field are quite small for the transmission coefficient, 
although the magnetization and the condensate wave function 
are easily changed by the magnetic field, as in Fig.~\ref{fig7}. 

A uniform linear Zeeman effect changes the spatial variation of the condensate wave function; 
however, the correspondence between the wave functions of excitations and the condensate wave functions 
still holds in the long-wavelength limit. 
As a result, the uniform magnetic field does not change the tunneling properties drastically, 
even in the unsaturated magnetization state. 
The tunneling properties depend on nonuniform effects, so that we expect that a nonuniform magnetic field changes the present results. 
We report this effect in a separate paper~\cite{WatabeKatoOhashi}. 

From the viewpoint of conducting experiments, 
the Zeeman effect is useful to study the tunneling problem of excitations in the spin-1 BEC. 
In the saturated magnetization state, the strong magnetic field can make 
spectra of three excitations (the Bogoliubov excitation, the transverse spin-wave mode, and the quadrupolar spin mode) separated well with an interval $g\mu_{\rm B}B$. 
An excitation in a BEC can be stimulated by the Bragg pulse, 
an advantage of which is that one can produce an excitation with a particular momentum $\hbar {\bf k}$ and a particular energy $\hbar \omega$ 
using two laser beams with different wave vectors and frequencies~\cite{Kozuma1999,Kurn1999,Steinhauer2002}. 
If the energy levels of excitations are well separated in the spin-1 BEC, 
it is easy to stimulate each mode selectively making use of this resonance scattering. 
If one performs experiments on this tunneling problem of the excitation in the spin-1 BEC, 
it is interesting to investigate the saturated magnetization state. 
The Bogoliubov excitation and the transverse spin wave show total transmission; 
but the quadrupolar spin mode shows total reflection against the potential wall. 
Rich scattering properties of excitations could be seen in the saturated magnetization state of a spin-1 spinor BEC.

\section{Summary}\label{Summary}

We determined the tunneling properties of excitations in saturated and unsaturated magnetization states of a spin-1 BEC. 
In the saturated magnetization state, 
we found that anomalous total transmission in the low-momentum limit occurs in a Bogoliubov mode and a transverse spin mode. 
However, a quadrupolar spin mode undergoes total reflection in the same limit. 
In the unsaturated magnetization state, however, 
all excitations show anomalous total transmission in the low momentum limit. 
When the anomalous tunneling phenomenon occurs, the phase shift is always zero. 

In order to investigate what determines total transmission and reflection in the low momentum limit, 
we focused on spectra of excitations and properties of wave functions. 
As for spectra of excitations, 
we have two points: the dispersion relation and the energy gap. 
As studied in Sec.~\ref{SecIII} for the saturated magnetization state, 
we found  total transmission, irrespective of the dispersion relation. 
The Bogoliubov excitation, whose energy $E$ is proportional to the momentum $p$ in the low momentum regime, 
and the transverse spin wave, with $E \propto p^{2}$, both show anomalous perfect tunneling in the low momentum limit. 

A key of anomalous tunneling was found in the energy gap. 
Indeed, we found that an excitation without an inherent energy gap shows total transmission in the long-wavelength limit; 
an excitation with an inherent energy gap shows total reflection. 
We use the term ``the inherent energy gap'' as the energy gap in the absence of spatially uniform external fields. 
For example, it is not the energy gap 
due to the linear Zeeman shift induced by the uniform magnetic field. 
In the saturated magnetization state, 
the quadrupolar spin mode which shows total reflection has the inherent energy gap; 
however, the Bogoliubov excitation and the spin-wave mode which show total transmission do not have the inherent energy gap.

Another key of anomalous tunneling was found in the properties of wave functions. 
We found that wave functions of excitations which show anomalous total transmission 
have the same forms as the condensate wave functions in the low-momentum limit. 
We note that the quadrupolar excitation showing total reflection has a wave function with zero amplitude in the long-wavelength limit, 
which is caused by the potential barrier. 
The spectra of excitations and the properties of wave functions suggest that the anomalous total transmission is strongly 
related to the existence of the zero modes in the absence of the uniform external field.

Table~\ref{tableII} summarizes the tunneling properties of excitations, their energy gap, and their wave functions in the low-momentum limit $p\rightarrow 0$. 
Here, $\Delta_{B = 0}$ denotes the inherent energy gap, 
and $\tilde{\phi}_{j}$ denotes the wave function of excitations with $j$ th component, 
$u_{j}$ and $v_{j}$ are the Bogoliubov-type wave functions given by $\tilde{\phi}_{j} \equiv u_{j} \exp{(-iEt/\hbar)} - v_{j}^{*} \exp{(iEt/\hbar)}$ with an energy $E$, and 
$\Phi_{j}$ denotes the condensate wave function with $j$ th component. 
In addition, 
$T$ and $\theta_{t}$, respectively, denote the transmission coefficient and the phase shift. 

\begin{table}[tbp]
\begin{center}
\caption{\label{tableII}
Relations between the tunneling properties of excitations, their energy gap, and their wave functions. 
}
Saturated magnetization state \quad  $(|{\bf m}| = 1)$  
\begin{tabular}{ c c c c c c c }
\hline\hline
\\[-10pt]
 \shortstack{Bogoliubov excitation} &&& 
\shortstack{Transverse spin \\ excitation} &&& \shortstack{Quadrupolar spin \\ excitation} 
\\
\hline
		           $\lim\limits_{p\rightarrow 0}T = 1$ &&& $\lim\limits_{p\rightarrow 0} T = 1$&&& $\lim\limits_{p\rightarrow 0}  T = 0$ 
		           \\ 
		           $\lim\limits_{p\rightarrow 0}\theta_{t} =0$ &&& $\lim\limits_{p\rightarrow 0} \theta_{t} =0$&&& $\lim\limits_{p\rightarrow 0} \theta_{t} \neq 0$ 
		           \\
$\Delta_{B = 0} = 0$ &&& $\Delta_{B = 0} = 0$ &&& $\Delta_{B = 0} \neq 0$
\\
		           $ \lim\limits_{p\rightarrow 0} (u_{+1}, v_{+1}) \propto (\Phi_{+1},\Phi_{+1})$ &&&
		           $\lim\limits_{p\rightarrow 0} \tilde{\phi}_{0} \propto \Phi_{+1}$ &&& 
		           $\lim\limits_{p\rightarrow 0} \tilde{\phi}_{-1} = 0$
		           \\ 
[5pt]
\hline\hline
\end{tabular}
\\[13pt]
Unsaturated magnetization state $(0\leq   |{\bf m}| < 1) $
\begin{tabular}{ c c c c c c c c c c  }
\hline\hline
\\[-10pt]
 \shortstack{Bogoliubov and longitudinal \\ spin excitation (out-of-phase)} &&& \shortstack{Bogoliubov and longitudinal \\ spin excitation (in-phase)} &&& \shortstack{Transverse spin \\ excitation}	
 \\
\hline
$\lim\limits_{p\rightarrow 0}T = 1$ &&& $\lim\limits_{p\rightarrow 0} T = 1$&&& $\lim\limits_{p\rightarrow 0} T = 1$
               \\
$\lim\limits_{p\rightarrow 0}\theta_{t} =0$ &&& $\lim\limits_{p\rightarrow 0} \theta_{t} =0$&&& $\lim\limits_{p\rightarrow 0} \theta_{t} =0$
 \\
$\Delta_{B = 0} = 0$ &&& $\Delta_{B = 0} = 0$ &&& $\Delta_{B = 0} = 0$   
		\\ 
		           \shortstack{$\lim\limits_{p\rightarrow 0} (u_{+1}, v_{+1},u_{-1}, v_{-1}) $ \\ $\propto (\Phi_{+1},\Phi_{+1}, \Phi_{-1},\Phi_{-1})$} 
		        &&& \shortstack{$\lim\limits_{p\rightarrow 0} (u_{+1}, v_{+1},u_{-1}, v_{-1}) $ \\ $\propto (-\Phi_{+1},-\Phi_{+1}, \Phi_{-1},\Phi_{-1})$}  
		        &&& \shortstack{$\lim\limits_{p\rightarrow 0} (u_{0}, v_{0})$ \\ $\propto (\Phi_{+1},\Phi_{-1})$}			
		\\
[5pt]
\hline\hline
\end{tabular}
\end{center}
\end{table}

We have some remaining issues. 
Here we studied the effects of the nonmagnetic potential barrier which couples only to the local density. 
It is unknown how a spin-dependent magnetic potential barrier affects the tunneling properties of excitations in spinor BECs. 
A spin-dependent magnetic potential barrier will drastically change the results of the nonmagnetic one. 
In the current carrying state of the scalar BEC, however, 
total transmission of the Bogoliubov excitation disappears at the critical current~\cite{Danshita2006}. 
In the current carrying state of the spinor BEC, it is also unknown what happens at the critical current for the tunneling problem. 
We report on these issues in a separate paper~\cite{WatabeKatoOhashi}.

\acknowledgements

We thank I. Danshita, D. Takahashi, T. Nikuni, M. Ueda, and H. Tsunetsugu for valuable discussions and comments. 
This work is supported by KAKENHI, Grant No. 21540352 from JSPS and 
Grant No. 20029007 from MEXT in Japan. 
SW acknowledges support from a Grant-in-Aid from JSPS (Grant No. 217751).

\appendix 
\section{EXCITATIONS}\label{AppendixA}

Here we summarize excitations in the spin-1 BEC on the basis of earlier studies~\cite{Ohmi1998,Ho1998}. 
We will derive eigenenergies and eigenvectors from Eqs.~(\ref{TBSpB7}) and (\ref{TBSpB8}) and discuss the properties of excitations. 

\subsection{Saturated magnetization state} 
In the uniform system, equations for excitations around the saturated magnetization state $(\Phi_{+1}, \Phi_{0}, \Phi_{-1})^{\rm T} = (\sqrt{n}, 0, 0)^{\rm T}$ are given by (\ref{FerroBogoEq1}), (\ref{FerroBogoEq2}), (\ref{Spin1ExcitationFerroEq0}), and (\ref{Spin1ExcitationFerroEqMinus}), with setting $\phi = 1$, as
\begin{align}
E 
\begin{pmatrix}
S_{+1} \\
G_{+1}
\end{pmatrix}
= &
\begin{pmatrix}
0 & - {\displaystyle \frac{\hbar^{2}}{2m} } \nabla^{2} + 2 c_{+} n \\
- {\displaystyle \frac{\hbar^{2}}{2m} } \nabla^{2} & 0
\end{pmatrix}
\begin{pmatrix}
S_{+1} \\
G_{+1}
\end{pmatrix}, 
\label{BogoFerro/2010/11/7/A1}
\\
E\tilde{\phi}_{0}
  = &
\left (
-\frac{\hbar^{2}}{2m}\nabla^{2}
+ 
g\mu_{\rm B}B
\right ) \tilde{\phi}_{0}, 
\label{BogoFerro/2010/11/7/A2}
\\
E\tilde{\phi}_{-1}
 = & 
\left (
-\frac{\hbar^{2}}{2m}\nabla^{2}
 -
2 c_{1} n \phi^{2}
+ 2g\mu_{\rm B}B
\right ) \tilde{\phi}_{-1}, 
\label{BogoFerro/2010/11/7/A3}
\end{align} 
where $S_{+1} \equiv u_{+1} + v_{+1}$, and $G_{+1} \equiv u_{+1} - v_{+1}$. 

For the spin component $+1$, the energy is given by 
$E = \sqrt{\varepsilon ( \varepsilon + 2c_{+}n ) }$, 
where $\varepsilon \equiv \hbar^{2}k^{2}/(2m)$, 
and we have the following solutions: 
\begin{align}
\begin{pmatrix}
S_{+1} \\
G_{+1}
\end{pmatrix} 
= 
\exp{(\pm i k x)}
\begin{pmatrix}
\alpha_{+1} \\
\beta_{+1}
\end{pmatrix}, 
\qquad 
\begin{pmatrix}
S_{+1} \\
G_{+1}
\end{pmatrix} 
= 
\exp{(\pm \kappa x)}
\begin{pmatrix}
\beta_{+1}
\\
- \alpha_{+1} 
\end{pmatrix}. 
\end{align}
Here, $k$ and $\kappa$ are given by 
\begin{align}
\hbar k &= \sqrt{ 2m 
\left [ \sqrt{(c_{+}n)^{2} + E^{2}} - c_{+} n \right ]}, 
\\
\hbar  \kappa &= \sqrt{ 2m 
\left [ \sqrt{(c_{+} n)^{2} + E^{2}} +  c_{+} n \right ] }. 
\end{align} 
Coefficients $\alpha_{+1}$ and $\beta_{+1}$, 
satisfying the normalization condition 
$(\alpha_{+1}^{*}\beta_{+1} + \alpha_{+1} \beta_{+1}^{*})/2 = 1$ (i.e., $|u_{+1}|^{2}-|v_{+1}|^{2}  = 1$), are given by 
\begin{align}
\alpha_{+1} \equiv \sqrt{ \frac{\sqrt{(c_{+} n)^{2} + E^{2}} + c_{+} n }{E} }, 
\\
\beta_{+1} \equiv \sqrt{ \frac{\sqrt{(c_{+} n)^{2} + E^{2}} - c_{+} n }{E} }. 
\end{align}

For the spin component $0$, 
the wave function $\tilde{\phi}_{0}$ is a solution to a Schr\"odinger-type equation (\ref{BogoFerro/2010/11/7/A2}) 
with the energy 
$E = \varepsilon + g\mu_{\rm B}B$. 
The solution in the uniform regime is given by $\tilde{\phi}_{0} = \exp{(\pm i k x )}$ with $\hbar k = \sqrt{ 2m (E - g\mu_{\rm B}B)}$. 

However, 
$\tilde{\phi}_{-1}$ is also a solution to another Schr\"odinger-type equation (\ref{BogoFerro/2010/11/7/A3}) 
with the energy 
$E = \varepsilon - 2 c_{1} n + 2g\mu_{\rm B}B$. 
The solution in the uniform regime is given by $\tilde{\phi}_{-1} = \exp{(\pm i k x )}$ with $\hbar k = \sqrt{ 2m \left [ E + 2 (c_{1}n-g\mu_{\rm B}B) \right ]}$. 

These fluctuations are related to the following modes~\cite{Ho1998,Ohmi1998}: 
$\tilde{\phi}_{+1}$ is associated with an ordinary Bogoliubov mode, 
which is composed of the density fluctuation $\delta {n} = \Phi_{+1} (\tilde{\phi}_{+1} +\tilde{\phi}_{+1}^{*} )$ 
and the phase fluctuation $\delta {\theta}_{+1} = (\tilde{\phi}_{+1} - \tilde{\phi}_{+1}^{*} )/ (2 i \Phi_{+1})$. 
$\tilde{\phi}_{0}$ is associated with the transverse spin wave mode $\delta {M}_{+} = \sqrt{2} \mu_{\rm B} \Phi_{+1}\tilde{\phi}_{0}$, 
and $\tilde{\phi}_{-1}$ is associated with the quadrupolar spin wave~\cite{Ho1998} (the longitudinal spin wave~\cite{Ohmi1998}) 
$\delta {M}_{-}^{2} =  2 \mu_{\rm B}^{2} \Phi_{+1} \tilde{\phi}_{-1}^{*}$.

\subsection{Unsaturated magnetization state}\label{Spin1SecPolarFM}

In this state, the configuration of order parameters is given by Eq. (\ref{TBSpB16}). 
In the uniform regime $V_{\rm ext}(|x|\rightarrow \infty) = 0$, 
we have the following equation: 
\begin{align}
E 
\begin{pmatrix}
S_{+1} \\
G_{+1} \\
S_{-1} \\
G_{-1} 
\end{pmatrix}
= 
\begin{pmatrix} 
0 & - {\displaystyle \frac{\hbar^{2}}{2m} }  \nabla^{2} +D_{+} & 0 & - D
\\
- {\displaystyle \frac{\hbar^{2}}{2m} }  \nabla^{2} & 0 & 0 & 0 
\\ 
0 & - D& 0 & - {\displaystyle \frac{\hbar^{2}}{2m} }  \nabla^{2} + D_{-}
\\ 
0 & 0 & - {\displaystyle \frac{\hbar^{2}}{2m} }  \nabla^{2} & 0
\end{pmatrix}
\begin{pmatrix}
S_{+1} \\
G_{+1} \\
S_{-1} \\
G_{-1} 
\end{pmatrix}, 
\end{align} 
where $D_{\pm } \equiv c_{+} n(1 \pm A)$ and $D \equiv c_{-} \sqrt{1-A^{2}}$ with $A$ being $g\mu_{\rm B} B/(c_{1}n)$.

The energy is given by 
\begin{align}
E_{(\pm)} &= \sqrt{\varepsilon ( \varepsilon + c_{+} n \pm C ) }, 
\end{align}
where $\varepsilon \equiv \hbar^{2}k^{2} / (2m)$ and $C \equiv \sqrt{(c_{-} n)^{2}  + 4c_{0}c_{1}n^{2}A^{2}}$. 
Solutions are given by 
\begin{align}
\begin{pmatrix}
S_{+1}^{(\pm)} \\
G_{+ 1}^{(\pm)} \\
S_{-1}^{(\pm)} \\
G_{-1}^{(\pm)}
\end{pmatrix} 
= 
\exp{(\pm i k_{(\pm)} x)}
\begin{pmatrix}
\alpha_{(\pm)} \\
\beta_{(\pm)} \\
\bar{\alpha}_{(\pm)} \\
\bar{\beta}_{(\pm)}
\end{pmatrix} , 
\qquad 
\begin{pmatrix}
S_{+1}^{(\pm)} \\
G_{+ 1}^{(\pm)} \\
S_{-1}^{(\pm)} \\
G_{-1}^{(\pm)} 
\end{pmatrix} 
= 
\exp{(\pm \kappa_{(\pm)} x)}
\begin{pmatrix}
- \beta_{(\pm)} \\
\alpha_{(\pm)} \\
-\bar{\beta}_{(\pm)} \\
\bar{\alpha}_{(\pm)} 
\end{pmatrix} , 
\end{align} 
where  the wavenumber $k_{(\pm)}$ and the diverging and converging rate $\kappa_{(\pm)}$ with the energy $E$ are given by 
\begin{align}
\hbar k_{(\pm)} = &\sqrt{ m 
\left [ 
\sqrt{(c_{+}n \pm C)^{2}+4E^{2}} - ( c_{+} n \pm C )
\right ] }, 
\label{ChapTB2-82}
\\
\hbar \kappa_{(\pm)} =  & \sqrt{ m 
\left [ 
\sqrt{( c_{+} n \pm C )^{2}+4E^{2}} + ( c_{+}n \pm C )
\right ] }. 
\label{ChapTB2-83}
\end{align}
Coefficients are obtained as 
\begin{align}
\begin{pmatrix}
\alpha_{(\pm)} \\
\beta_{(\pm)} \\
\bar{\alpha}_{(\pm)} \\
\bar{\beta}_{(\pm)}
\end{pmatrix} 
= 
\begin{pmatrix}
\displaystyle{
\mp
\sqrt{
\frac{2E (C\pm c_{+} nA)}
{C [ \sqrt{(c_{+} n \pm C)^{2} + 4E^{2}} - ( c_{+} n \pm C ) ] }
}
}
\\
\\
\displaystyle{
\mp
\sqrt{
\frac{2E ( C\pm c_{+} nA ) }
{C [ \sqrt{(c_{+} n \pm C)^{2} + 4E^{2}} + (c_{+} n \pm C) ] }
}
}
\\
\\
\displaystyle{
\sqrt{
\frac{2E ( C\mp c_{+}nA ) }
{C [ \sqrt{(c_{+}n \pm C)^{2} + 4E^{2}} - (c_{+} n \pm C ) ]}
}
}
\\
\\
\displaystyle{
\sqrt{
\frac{2E (C\mp c_{+} nA ) }
{C [ \sqrt{(c_{+} n \pm C) ^{2} + 4E^{2}} + ( c_{+} n \pm C) ] }
}
}
\end{pmatrix},  
\end{align} 
which satisfy the normalization condition 
$
{\rm Re} [\alpha_{(\pm)} \beta_{(\pm)} + \bar{\alpha}_{(\pm)} \bar{\beta}_{(\pm)}] = 2
$, corresponding to $\sum\limits_{i=\pm1}(|u_{i}|^{2}-|v_{i}|^{2})  = 2$.

However, the equation of the spin component $0$ is given by 
\begin{align}
E 
\begin{pmatrix}
u_{0} \\ 
v_{0}
\end{pmatrix}
= 
\begin{pmatrix}
- {\displaystyle \frac{\hbar^{2}}{2m} } \nabla^{2} + c_{1} n & c_{1} n \sqrt{1-A^{2}} \\
- c_{1} n \sqrt{1-A^{2}} & 
- \left ( - {\displaystyle \frac{\hbar^{2}}{2m} } \nabla^{2} + c_{1} n \right ) 
\end{pmatrix}
\begin{pmatrix}
u_{0} \\ 
v_{0}
\end{pmatrix}.  
\end{align}   
The energy of this mode is obtained as 
\begin{align}
E & = \sqrt{\varepsilon (\varepsilon + 2c_{1} n)+ (g\mu_{\rm B}B)^{2}}. 
\end{align}
Its eigenvectors can be written as 
\begin{align}
\begin{pmatrix}
u_{0} \\ 
v_{0}
\end{pmatrix} 
= 
\exp{(\pm ikx)}
\begin{pmatrix}
\alpha_{0} \\
\beta_{0} 
\end{pmatrix},  
\qquad 
\begin{pmatrix}
u_{0} \\ 
v_{0}
\end{pmatrix}
= 
\exp{(\pm \kappa x)}
\begin{pmatrix} 
- \beta_{0} \\ 
\alpha_{0} 
\end{pmatrix},
\end{align}  
where the wavenumber $k$ and the diverging and converging rate $\kappa$ are given by 
\begin{align}
\hbar k = & \sqrt{ 2m 
\left [ 
\sqrt{( c_{1}n )^{2} (1-A^{2}) + E^{2} } -c_{1}n
\right ]}, 
\label{ChapTB2-88}
\\ 
\hbar \kappa = & \sqrt{ 2m  
\left [ 
\sqrt{( c_{1}n )^{2} (1-A^{2}) + E^{2} } + c_{1}n
\right ]}. 
\label{ChapTB2-89}
\end{align}
Here, $\alpha_{0}$ and $\beta_{0}$ are given by 
\begin{align}
\begin{pmatrix}
\alpha_{0} \\
\beta_{0} 
\end{pmatrix}
= 
\begin{pmatrix}
\displaystyle{
\sqrt{
\frac{
\sqrt{ (c_{1}n)^{2} {(1-A^{2})} + E^{2}} + E
}
{2E}
}
}
\\
-
\displaystyle{
\sqrt{
\frac{
\sqrt{ (c_{1}n)^{2} {(1-A^{2})} + E^{2}} - E
}
{2E}
}
}
\end{pmatrix}. 
\end{align}  

Eigenmodes with $E_{(+)}$ and $E_{(-)}$ respectively correspond to out-of-phase and in-phase modes between the spin components $\pm 1$, 
since signs of $S_{+1}^{(+)}$ and $S_{-1}^{(+)}$ (and also $G_{+1}^{(+)}$ and $G_{-1}^{(+)}$ ) are opposite, 
and signs of $S_{+1}^{(-)}$ and $S_{-1}^{(-)}$ (and also $G_{+1}^{(-)}$ and $G_{-1}^{(-)}$ ) are the same. 
In Ref.~\cite{Ohmi1998}, 
these modes were termed the collective mode that couples with the number density and the longitudinal spin density. 
Specifically in the low energy limit, these excitations can be regarded as phase modes of condensates like the Bogoliubov mode. 
We thus term the mode with $E_{(+)}$ $[E_{(-)}]$, 
``Bogoliubov and longitudinal spin excitation [out-of-phase (in-phase)]''. 
The other mode $\tilde{\phi}_{0}$ is associated with the transverse spin wave mode 
$\delta {M}_{+} = \delta {M}_{-}^{*} = \sqrt{2}\mu_{\rm B}(\Phi_{+1}\tilde{\phi}_{0} + \Phi_{-1}\tilde{\phi}_{0}^{*} )$~\cite{Ohmi1998}.

\section{INTEGRABLE CASE}\label{AppendixC}

As seen in Eq.~(\ref{GPPolarFM+1}) for 
the unsaturated magnetization state, 
the condition $c_{0} = c_{1}$ makes the GP-type equation of the spin components $\pm 1$ decoupled. 
It also holds for the Bogoliubov-type equation (\ref{BogoliubovPolarFMpm1}). 
The condition $c_{0} = c_{1}$ corresponds to the integrable condition found by Ieda {\it et al.}~\cite{Ieda2004, Ieda2004II, Uchiyama2006}. 
We shall consider the special case $c_{0} = c_{2} (\equiv \tilde{c})$.

Using this condition, 
the GP-type equation is given by 
\begin{align}
0 & = 
\left (
-\frac{\hbar^{2}}{2m}\nabla^{2}
+ V_{\rm ext} -\tilde{c} n \mp g\mu_{\rm B}B 
\right ) \Phi_{\pm 1}
+ 
2\tilde{c}\Phi_{\pm1}^{3}. 
\label{eq:B1}
\end{align} 
We assume that the configuration is given by  
\begin{align}
( \Phi_{+1}, \Phi_{0}, \Phi_{-1} )^{\rm T}
= 
(-\sqrt{(1+A)n/2} \phi_{+1} (x), 0, \sqrt{(1-A)n/2} \phi_{-1}(x))^{\rm T}, 
\end{align} 
where $\phi_{\pm 1} (|x| \rightarrow \infty) = 1$. 
Substituting it into the GP-type equation, we obtain the equation 
\begin{align}
0 & = 
\left [ 
-\frac{\hbar^{2}}{2m}\nabla^{2}
+ V_{\rm ext} (x)
+ \tilde{c}n(1\pm  g\mu_{\rm B}B) 
(\phi_{\pm 1}^{2} (x) -1)
\right ]
\phi_{\pm 1} (x). 
\end{align} 
This is the same form as the equation of the scalar BEC, and hence it is easy to solve. 

Far from the potential barrier, 
however, we have the equation of the excitation given by 
\begin{align}
E_{\pm} 
\begin{pmatrix}
S_{\pm 1} \\
G_{\pm 1} 
\end{pmatrix}
= 
\begin{pmatrix}
0 & - {\displaystyle \frac{\hbar^{2}}{2m}} \nabla^{2} + 2 (\tilde{c} \pm  g\mu_{\rm B}B) \\
- {\displaystyle \frac{\hbar^{2}}{2m}} \nabla^{2} & 0 
\end{pmatrix}
\begin{pmatrix}
S_{\pm 1} \\
G_{\pm 1} 
\end{pmatrix}. 
\end{align} 
The energy is obtained as 
\begin{align}
E_{\pm} &= \sqrt{\varepsilon [\varepsilon + 2(\tilde{c}n \pm  g\mu_{\rm B}B)]}, 
\end{align}
and solutions are given by 
\begin{align}
\begin{pmatrix}
S_{\pm 1} \\
G_{\pm 1} 
\end{pmatrix}
= 
\exp{(\pm ikx)} 
\begin{pmatrix}
\alpha_{\pm1} \\
\beta_{\pm 1} 
\end{pmatrix}, 
\qquad 
\begin{pmatrix}
S_{\pm 1} \\
G_{\pm 1} 
\end{pmatrix}
= 
\exp{(\pm \kappa x)} 
\begin{pmatrix}
\beta_{\pm 1}  \\
- \alpha_{\pm1}
\end{pmatrix}, 
\end{align} 
where 
\begin{align}
\hbar k_{\pm} &= \sqrt{ 2m 
\left [ 
\sqrt{(\tilde{c}n \pm  g\mu_{\rm B}B)^{2} + E_{\pm}^{2}}- (\tilde{c}n \pm  g\mu_{\rm B}B)
\right ] }, 
\\
\hbar \kappa_{\pm} &= \sqrt{ 2m 
\left [ 
\sqrt{(\tilde{c}n \pm  g\mu_{\rm B}B)^{2} + E_{\pm}^{2}} + (\tilde{c}n \pm  g\mu_{\rm B}B)
\right ]}, 
\end{align}
and 
\begin{align}
\alpha_{\pm1} = & 
\sqrt{ 
\frac{\sqrt{(\tilde{c}n \pm  g\mu_{\rm B}B)^{2} + E_{\pm}^{2}}+ (\tilde{c}n \pm  g\mu_{\rm B}B)}{E_{\pm}}
}, 
\\ 
\beta_{\pm 1} = & 
\sqrt{
\frac{\sqrt{(\tilde{c}n \pm  g\mu_{\rm B}B )^{2} + E_{\pm}^{2}} - (\tilde{c}n \pm  g\mu_{\rm B}B )}{E_{\pm}}
}. 
\end{align}

We obtain the transmission coefficient by solving the following equations: 
\begin{align}
E G_{\pm 1} 
&= 
\left [
-\frac{\hbar^{2}}{2m}\nabla^{2}
+ V_{\rm ext} 
- \tilde{c} n \mp  g\mu_{\rm B}B  
+ 2\tilde{c} \Phi_{\pm1}^{2} 
\right ] 
S_{\pm 1}, 
\label{eq:B11}
\\
E S_{\pm 1} 
&= 
\left [
-\frac{\hbar^{2}}{2m}\nabla^{2}
+ V_{\rm ext} 
- \tilde{c} n \mp  g\mu_{\rm B}B  
+ 6 \tilde{c} \Phi_{\pm1}^{2} 
\right ] 
G_{\pm1}, 
\end{align}
imposing the boundary conditions 
\begin{align}
\begin{pmatrix}
S_{\pm} \\
G_{\pm} 
\end{pmatrix} 
= & 
e^{ik_{\pm} x}
\begin{pmatrix}
\alpha_{\pm} \\
\beta_{\pm} 
\end{pmatrix} 
+ 
r_{\pm}
e^{-ik_{\pm} x}
\begin{pmatrix}
\alpha_{\pm} \\
\beta_{\pm}
\end{pmatrix}  
+ 
a_{\pm}
e^{\kappa_{\pm}  x}
\begin{pmatrix}
\beta_{\pm} \\
- \alpha_{\pm}
\end{pmatrix} 
\qquad & 
{\rm for} \, x \rightarrow - \infty, 
\\
\begin{pmatrix}
S_{\pm} \\
G_{\pm}
\end{pmatrix} 
= & 
t_{\pm} 
e^{ik_{\pm}  x}
\begin{pmatrix}
\alpha_{\pm} \\
\beta_{\pm}
\end{pmatrix}  
+ 
b_{\pm} 
e^{-\kappa_{\pm}  x}
\begin{pmatrix}
\beta_{\pm} \\
- \alpha_{\pm}
\end{pmatrix} 
\qquad &
{\rm for} \, x \rightarrow + \infty. 
\end{align}

\begin{figure}[tbp]
\begin{center}
\includegraphics[width=8.5cm]{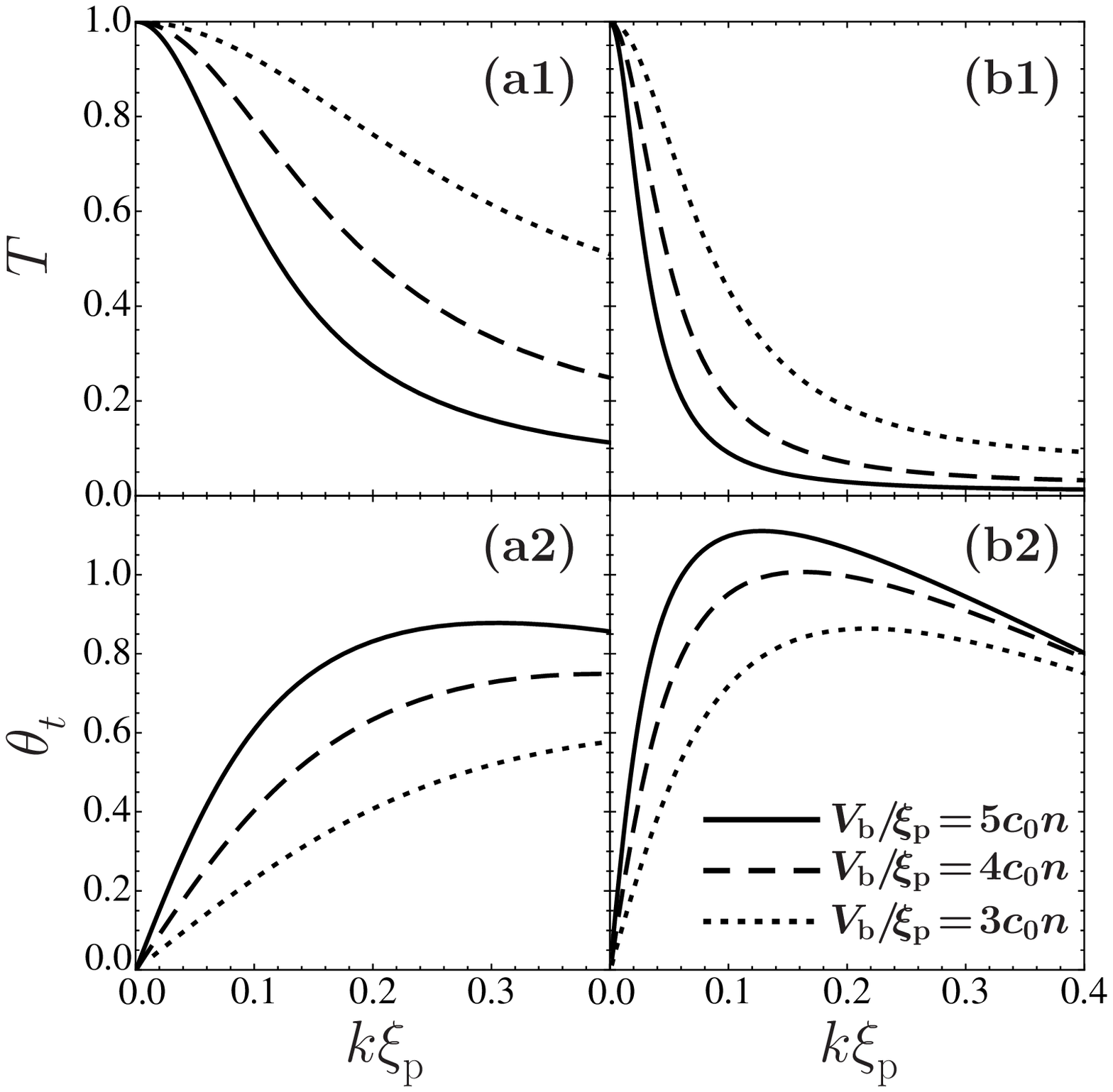}
\end{center}
\caption{ 
Tunneling properties of excitations in the integrable case $(c_{0} = c_{1})$. 
(a1) Transmission coefficients and 
(a2) phase shifts of the spin component $+1$. 
(b1) Transmission coefficients and (b2) phase shifts of the spin component $-1$. 
We use the $\delta$-function potential barrier $V_{\rm ext} (x) = V_{\rm b} \delta (x)$. 
We use parameters $a_{0} : a_{2} = 46 : 52$, following Ref.~\cite{Ho1998}. 
We also use $g\mu_{\rm B} B = 0.5 \tilde{c} n$.  }
\label{fig16}
\end{figure}

By comparing (\ref{eq:B1}) and (\ref{eq:B11}) for the limit $E\rightarrow 0$, 
we find that the wave function $S_{\pm 1}$ in the long-wavelength limit 
has a solution corresponding to the condensate wave function $\Phi_{\pm 1}$. 
We expect perfect transmission of both modes in the long wavelength limit. 
Figure~\ref{fig16} shows the results of the tunneling problems for the spin components $\pm 1$. 
We find the total transmission in the long-wavelength limit, 
and find that each phase shift goes to zero in this limit. 
In the integrable case $c_{0} = c_{1}$, 
we conclude that 
the fluctuations of different spin components are decoupled, 
and each excitation shows the total transmission in the long wavelength limit. 

\begin{thebibliography}{99}



\bibitem{Kovrizhin2001}
D. L. Kovrizhin, {Phys. Lett. A} {\bf 287}, 392 (2001). 

\bibitem{Kagan2003}
Yu. Kagan, D. L. Kovrizhin, and L. A. Maksimov, {Phys. Rev. Lett.} {\bf 90}, 130402 (2003). 

\bibitem{Danshita2006}
I. Danshita, N. Yokoshi, and S. Kurihara, {New J. Phys.} {\bf 8}, 44 (2006). 

\bibitem{FujitaMThesis}
A. Fujita, Master's thesis The University of Tokyo, 2007.

\bibitem{FujitaUnpublished}
A. Fujita, and Y. Kato (unpublished). 

\bibitem{Kato2007}
Y. Kato, H. Nishiwaki, and A. Fujita, {J. Phys. Soc. Jpn.} {\bf 77}, 013602 (2008). 

\bibitem{Watabe2008}
S. Watabe and Y. Kato, {Phys. Rev. A} {\bf 78}, 063611 (2008). 

\bibitem{Tsuchiya2008} S. Tsuchiya and Y. Ohashi, Phys. Rev. A {\bf 78}, 013628 (2008). 

\bibitem{Ohashi2008} Y. Ohashi, and S. Tsuchiya, Phys. Rev. A {\bf 78}, 043601 (2008). 

\bibitem{Watabe2009RefleRefra}
S. Watabe and Y. Kato, {Journal of Physics : Conference Series}, {\bf 150}, 032119 (2009). 

\bibitem{Takahashi2009}
D. Takahashi, and Y. Kato, {J. Phys. Soc. Jpn.} {\bf 78}, 023001 (2009). 

\bibitem{Tsuchiya2009}
S. Tsuchiya and Y. Ohashi, Phys. Rev. A {\bf 79}, 063619 (2009). 

\bibitem{WatabeKato2009}
S. Watabe and Y. Kato, {J. Low Temp. Phys.} {\bf 158}, 23 (2010). 

\bibitem{Takahashi2010}
D. Takahashi, and Y. Kato, {J. Low Temp. Phys.} {\bf 158}, 65 (2010). 

\bibitem{WatabeKatoLett}
S. Watabe and Y. Kato, arXiv:1012.5618. 

\bibitem{Stamper-Kurn1998}
D. M. Stamper-Kurn, M. R. Andrews, A. P. Chikkatur, S. Inouye, H.-J. Miesner, J. Stenger, and W. Ketterle, 
Phys. Rev. Lett. {\bf 80}, 2027 (1998). 

\bibitem{Stenger1998}
J. Stenger, S. Inouye, D. M. Stamper-Kurn, H.-J. Miesner, A. P. Chikkatur, and W. Ketterle, 
Nature (London) {\bf 396}, 345 (1998). 

\bibitem{Ohmi1998}
T. Ohmi, and K. Machida, {J. Phys. Soc. Jpn.} {\bf 67}, 1822 (1998). 

\bibitem{Ho1998}
Tin-Lun Ho, {Phys. Rev. Lett.} {\bf 81}, 742 (1998). 

\bibitem{Demokritov2004}
S. O. Demokritov, A. A. Serga, A. Andr\'e, V. E. Demidov, M. P. Kostylev, B. Hillebrands, 
and A. N. Slavin, 
{Phys. Rev. Lett.} {\bf 93}, 047201 (2004). 

\bibitem{Hansen2007}
Ulf-Hendrik Hansen, Marius Gatzen, Vladislav E. Demidov, and Sergej O. Demokritov, 
{Phys. Rev. Lett.} {\bf 99}, 127204 (2007). 

\bibitem{Nistazakis2007}
H. E. Nistazakis, D. J. Frantzeskakis, P. G. Kevrekidis, B. A. Malomed, R. Carretero-Gonz\'alez, and A. R. Bishop, {Phys. Rev. A} {\bf 76}, 063603 (2007). 

\bibitem{NoteMode}
For example, in the saturated magnetization state, 
when we assume that the incident mode $\sigma = {\rm I}$ is the Bogoliubov mode, 
we can label the two other modes (i.e., the transverse spin mode and the quadrupolar spin mode) which may appear as reflected and transmitted waves as II and III. 

\bibitem{Numerical}
We found that the transmission and reflection coefficients of the opposite type of the excitation 
are at most ${\mathcal O} (10^{-3})$ in our numerical calculation. 

\bibitem{Fetter1972} 
A. L. Fetter, {Ann. Phys.} {\bf 70}, 67 (1972). 

\bibitem{WatabeKatoOhashi}
S. Watabe, Y. Kato, and Y. Ohashi, Phys. Rev. A {\bf 83}, 033627 (2011). 

\bibitem{Hohenberg1965}
P. C. Hohenberg, and P. C. Martin, {Ann. Phys. (N. Y.)} {\bf 34}, 291 (1965). 

\bibitem{Mimic}
For example, 
we confirm the existence of type I in the tunneling problem of a quantum particle obeying the Schr\"odinger equation 
against the following barrier: [$V(x) = 0.6 V_{\rm b}$ for $|x|< l_{\rm s}$, $V(x) = -0.4V_{\rm b}$ for $ l_{\rm s} \leq |x| < 2 l_{\rm s}$, and  $V (x) = 0$ for $2 l_{\rm s} \leq |x|$], 
where $ l_{\rm s} \equiv \hbar/\sqrt{m V_{\rm b} }$. 
Type II can be confirmed against the following barrier: [$V (x) = 0.5 V_{\rm b}$ for $|x|< l_{\rm s}$, $V (x) = -0.5V_{\rm b}$ for $ l_{\rm s}\leq |x| < 2  l_{\rm s}$, 
and $V (x) = 0$ for $2 l_{\rm s} \leq |x|$]. 

\bibitem{Kozuma1999}
M. Kozuma, L. Deng, E. W. Hagley, J. Wen, R. Lutwak, K. Helmerson, S. L. Rolston, and W. D. Phillips, Phys. Rev. Lett. {\bf 82}, 871 (1999). 

\bibitem{Kurn1999}
D. M. Stamper-Kurn, A. P. Chikkatur, A. G\"orlitz, S. Inouye, S. Gupta, D. E. Pritchard, and W. Ketterle, Phys. Rev. Lett. {\bf 83}, 2876 (1999). 

\bibitem{Steinhauer2002}
J. Steinhauer, R. Ozeri, N. Katz, and N. Davidson, Phys. Rev. Lett. {\bf 88}, 120407 (2002). 

\bibitem{Ieda2004}
J. Ieda, T. Miyakawa, and M. Wadati, 
{Phys. Rev. Lett.} {\bf 93}, 194102 (2004). 

\bibitem{Ieda2004II}
J. Ieda, T. Miyakawa, and M. Wadati, {J. Phys. Soc. Jpn.} {\bf 73}, 2996 (2004). 

\bibitem{Uchiyama2006}
M. Uchiyama, J. Ieda, and M. Wadati, 
{J. Phys. Soc. Jpn.} {\bf 75}, 064002 (2006). 

\end{thebibliography}
\end{document}